\documentclass[aps,prl,reprint,twocolumn,superscriptaddress,nofootinbib]{revtex4-2}

\usepackage{amsmath,amssymb}
\usepackage{bm}
\usepackage{graphicx}
\usepackage{xcolor}
\usepackage{booktabs}
\usepackage{microtype}
\usepackage{hyperref}
\hypersetup{hidelinks}

\newcommand{\bk}{\bm{k}}
\newcommand{\bq}{\bm{q}}
\newcommand{\bd}{\bm{d}}
\newcommand{\bsigma}{\bm{\sigma}}
\newcommand{\bs}{\bm{s}}
\newcommand{\bb}{\bm{b}}
\newcommand{\bbeta}{\bm{\beta}}
\newcommand{\Tr}{\operatorname{Tr}}
\newcommand{\tr}{\operatorname{tr}}
\newcommand{\dd}{\mathrm{d}}
\newcommand{\orb}{\mathrm{orb}}
\newcommand{\so}{\mathrm{so}}
\newcommand{\FS}{\mathrm{FS}}

\begin{document}

\title{Orbital-geometric spin response: inter-band coherence versus orbital-to-spin conversion}

\author{Dimitrie Culcer}
\affiliation{School of Physics, The University of New South Wales, Sydney 2052, Australia}
\date{}

\begin{abstract}
Orbitronics seeks to convert orbital angular momentum (OAM) into spin, but its microscopic dynamics is not understood. Here I show that, at moderate spin-orbit coupling, an electrically induced spin density contains a direct term and an \textit{orbital-geometric spin response} (OGSR) due to inter-band coherence of the spinless parent Hamiltonian---related to OAM but not descended from it. In massive Dirac fermions the OGSR survives as the net OAM vanishes, and generates \textit{counterflow}, opposing the direct response. These results provide guidelines for optimising orbitronic systems.
\end{abstract}
\date{\today}
\maketitle

\textit{Introduction}. Understanding the connection between Bloch electrons' orbital dynamics and their spin is a central problem in orbitronics, a field that focuses on generating non-equilibrium orbital angular momentum (OAM) densities to manipulate an adjacent magnetization~\cite{Go2021,Atencia2024,Jo2024,CysneReview2025,Fukami2026}. Theoretical and experimental work have established strong orbital responses in a broad range of structures~\cite{Bernevig2005,Kontani2009,Go2018,Jo2018,Canonico2020,Ding2020,Cysne2021,Lee2021Conversion,Lee2021OrbitalTorque,Ding2022,Cysne2022,Pezo2022,Hayashi2023,Choi2023,Seifert2023,Lyalin2023,Cullen2025,Cullen2026Holes}. Yet OAM does not couple directly to the magnetization. Instead, a steady-state OAM density accompanied by spin-orbit coupling (SOC) is expected to result in a spin density, which ultimately exerts a torque on the magnetization~\cite{Ding2020,Lee2021Conversion,Hayashi2023,Pezo2025Pumping,Pezo2025CoAl,Go2025Pumping}. This phenomenon is commonly described experimentally as orbital-to-spin conversion~\cite{Chen2026Switching}, but its microscopic basis remains poorly understood; an atom-centred continuity framework has nevertheless clarified the local transfer and interconversion of spin and orbital angular momentum~\cite{Go2020}. Crucially, it remains unknown whether the OAM response, as understood by the modern theory of orbital magnetization~\cite{Vanderbilt2018}, is an intermediate variable controlling spin generation, and whether a sequential picture of the form electric field $\rightarrow$ OAM $\rightarrow$ spin is justified. The central question is therefore: \textit{How much of the non-equilibrium spin density is due to orbital dynamics, and how much of it can be ascribed to the OAM?}

The difficulty is fundamental. Firstly, within the modern theory, an electron's equilibrium orbital moment is envisaged as the self rotation of a wave packet travelling through the lattice~\cite{Resta1998,Xiao2005,Thonhauser2005,Ceresoli2006,Shi2007,Xiao2010,Bianco2013,Vanderbilt2018,Souza2008,Lopez2012,Resta2020,Vignale2026}. The rotation is associated with a quantum geometric quantity stemming from inter-band dynamics, while the modern-theory formula does not determine the microscopic spin-orbit coupling through which that orbital motion generates spin. In fact, spin-orbit coupling in the band structure typically enhances inter-band coherence and itself acts as a source of OAM~\cite{Unzelmann2020,CullenGe2026,Cullen2026Holes}.
Thus, from the point of view of the modern theory, neither OAM nor orbital dynamics can simply be assumed to convert into spin. Secondly, the orbital magnetoelectric response contains intra- and inter-band density-matrix contributions~\cite{Yoda2015,Yoda2018,Hara2020,Zhong2016,Salemi2019,He2020,Johansson2021,Ado2024,CullenOME2026}, while disorder and boundary scattering further complicate OAM and spin dynamics~\cite{Aronov1989,Edelstein1990,Valet1993,Sinova2004,Shi2006,Nagaosa2010,Scaramucci2012,Raimondi2012,Oyarzun2016,Culcer2017,Otani2017,Shitade2018,Manchon2019,LeeEnsslin2020,Veneri2022,LeeCasanova2022,Mercaldo2023,YangYan2023,AdoPosition2024,Atencia2024,Canonico2024,Liu2025,Sun2025,Voss2025,Veneri2025,Aase2025,AdoMagnetic2026,Cysne2026}. No framework has established how these channels combine within the modern theory.

In this work I develop a general method for identifying the quantum geometric component of the non-equilibrium spin response stemming explicitly from orbital dynamics, and for testing whether the complete OAM tensor reappears with the same occupations and energy weights. Starting from a spinless parent Hamiltonian, which has no spin-orbit coupling but exhibits strong inter-band dynamics, I consider the effect of adding spin-orbit coupling, first generically and then perturbatively. The contribution to the spin density at a given wave vector ${\bm k}$ due to inter-band coherence in the parent Hamiltonian is termed the \textit{orbital-geometric spin response} (OGSR). In the weak-SOC, momentum-independent models below it is distinguished from the direct term, which stems from the SOC correction to the band energies. Establishing the exact form of the OGSR and its relationship to OAM requires an explicit model, and massive Dirac fermions provide an ideal test case. An untilted massive Dirac cone possesses an equilibrium ${\bm k}$-resolved OAM but no orbital magneto-electric effect, whereas a tilt generates non-equilibrium OAM, allowing the equilibrium and non-equilibrium contributions to be tracked separately. Importantly for untilted massive Dirac fermions, whose non-equilibrium spin density points in the plane, (i) the OGSR is nonzero while the net OAM density is zero; (ii) the OGSR opposes the direct contribution, generating an \textit{orbital counterflow} that cancels $93$--$94\%$ of the direct response for representative transition-metal dichalcogenide parameters; (iii) to first order in SOC the OGSR cannot be identified with either the equilibrium or non-equilibrium OAM, and (iv) the first identifiable OAM-derived terms appear at second order in SOC. On the other hand, the tilt allows an out-of-plane spin density which can be expressed as a function of the non-equilibrium OAM. These results show that spin can be generated via orbital-geometric channels without the modern-theory OAM response acting as an intermediate dynamical variable. Thus inter-band coherence is the fundamental starting point in dynamics associated with the OAM \cite{IorshTitovKerr2026}, and the distinction provides a design principle for optimising orbitronic torque devices.

\textit{Hamiltonian}. The full Hamiltonian is $H=H_{\orb}+H_{\so}$, where $H_{\orb}$ is the spinless parent Hamiltonian and $H_{\so}$ represents the SOC. This is a physical rather than an arbitrary partition: $H_{\orb}$ is obtained by adiabatically removing the microscopic SOC while retaining the same orbital basis and all spin-independent terms. In a first-principles implementation it is the scalar-relativistic Wannier Hamiltonian, while $H_{\so}$ is the difference from the fully relativistic Hamiltonian in the same Wannier gauge. They are resolved in terms of spectral projector operators as 
$H_{\orb}=\sum_A\varepsilon_AP_A$ and
$H = \sum_NE_N\Pi_N$ respectively. The projectors $P_A$ define orbital manifolds and include the identity in real spin, while $\Pi_N$ describe the exact SOC-coupled bands. Both may have rank greater than one. Differentiating the parent eigenvalue equation gives
$v_j^{\orb}=\hbar^{-1}\sum_A(\partial_j\varepsilon_A)P_A+(i/\hbar)\sum_{A\ne B}(\varepsilon_A-\varepsilon_B){\cal R}_{AB}^{0,j}$,
where ${\cal R}_{AB}^{0,j}=iP_A\partial_jP_B$ is the projector form of the Berry connection matrix. I denote the two individual terms in this decomposition by $v_j^{\orb,\parallel}$ and $v_j^{\orb,\perp}$. They are separately Hermitian and gauge covariant within every parent manifold. A momentum-dependent SOC produces two further blocks, which are retained in the exact response but do not obey this universal identity; their treatment is given in the Supplemental Material~\cite{Supplement}.
Explicitly, for $\chi=\orb,\so$ I define
$v_j^{\chi,\parallel}=\sum_AP_Av_j^\chi P_A$ and $v_j^{\chi,\perp}=\sum_{A\ne B}P_Av_j^\chi P_B$,
so that the physical velocity is the sum of the four $(\chi,\parallel/\perp)$ blocks. Only $v_j^{\orb,\perp}$ has the universal Berry-connection form given above; $v_j^{\so}=\hbar^{-1}\partial_jH_{\so}$ depends on the microscopic SOC, which is kept separate for classification purposes. For this fixed physical parent the decomposition is additive and basis independent inside every composite manifold; it is not intended to be invariant under redefining the parent itself.

For the remainder of this work I specialize to isolated nondegenerate orbital bands $n$ and exact SOC bands $N$. The orbital part of $P_n=|u_n\rangle\langle u_n|\otimes 1_s$ is rank one, real spin remains an explicit two-dimensional space, and $\Pi_N=|\Psi_N\rangle\langle\Psi_N|$. Thus ${\cal R}_{nm}^{j}=i\langle u_n|\partial_j u_m\rangle$ is the ordinary parent-interband Berry connection. At crossings generated by removing SOC, the connected bands must be grouped into a composite parent manifold. The projector construction suitable for arbitrary Wannier Hamiltonians ~\cite{Marzari2012,Mera2022,Mitscherling2025,VermaQueiroz2025,Go2024,Lee2026} is derived in the Supplement~\cite{Supplement}.

\textit{Orbital-geometric spin response}. The leading weak-disorder density matrix in an electric field is diagonal in the exact-band basis, and in the relaxation-time approximation (RTA) it takes the form
\begin{equation}
(\rho_E^{d})_{NN}=e\tau E_jv_{jN}(-f_N'),
\label{eq:diagonal_dm}
\end{equation}
where the electron charge is $-e$, $\tau$ is the transport relaxation time, $v_{jN}=\langle\Psi_N|v_j|\Psi_N\rangle$, $f_N=f(E_N)$ is the equilibrium Fermi function, and $f_N'=\partial f(E_N)/\partial E_N$. Repeated Cartesian indices are summed and $\int_{\bk}=\int\dd^dk/(2\pi)^d$ in $d$ spatial dimensions. This leading band-diagonal treatment excludes scattering-induced interband coherence and vertex corrections. The resulting Fermi-surface spin density is
\begin{align}
S_a^{\FS}&=e\tau E_j\sum_X C_{aj}^{X},
\\
C_{aj}^{X}&=\sum_N\int_{\bk}(-f_N')s_{aN}v_{jN}^{X},
\label{eq:nondeg_response}
\end{align}
with $s_a$ a dimensionless Pauli spin matrix, $s_{aN}=\langle\Psi_N|s_a|\Psi_N\rangle$, $v_{jN}^{X}=\langle\Psi_N|v_j^{X}|\Psi_N\rangle$, and $X\in\mathcal X\equiv\{(\orb,\parallel),(\orb,\perp),(\so,\parallel),(\so,\perp)\}$. Thus $S_a$ is the Pauli-spin polarisation density; multiplication by $\hbar/2$ gives the spin-angular-momentum density. The Fermi-surface OGSR is $C_{aj}^{\rm OG}\equiv C_{aj}^{\orb,\perp}$, which couples the spin of each exact eigenstate to the parent-interband velocity; the other three blocks complete the exact response. For the momentum-independent SOC used below, $v_j^{\so}=0$ and the response reduces to $C^{\orb,\parallel}+C^{\rm OG}$. I then write these as $C^{\parallel}+C^{\perp}$ and refer to them as the direct and OGSR contributions, respectively; at weak SOC the former is generated by the SOC correction to the projected band energies. When $C^{\perp}$ opposes $C^{\parallel}$ it represents \emph{counterflow}. Both terms arise from the same band-diagonal, lifetime-dependent $\rho_E^{d}$. The Fermi-surface OGSR reflects parent-interband coherence already present in each SOC-coupled eigenstate.

The electric field also generates exact-band spin coherence through the off-diagonal density matrix $\rho_E^{od}$ of the SOC Hamiltonian ~\cite{Culcer2017,ValetRaimondi2025}. For nondegenerate exact bands, the clean band-intrinsic Kubo response~\cite{Kubo1957} can be decomposed on the same footing,
\begin{align}
S_a^{\rm int}&=e\hbar E_j\sum_X\widetilde C_{aj}^{X},
\nonumber\\
\widetilde C_{aj}^X &=2\sum_{N<M}\int_{\bk}
\frac{f_N-f_M}{(E_N-E_M)^2}
\operatorname{Im}[s^a_{MN}v^{X,j}_{NM}],
\label{eq:intrinsic_general}
\end{align}
where $s^a_{MN}=\langle\Psi_M|s_a|\Psi_N\rangle$, $v^{X,j}_{NM}=\langle\Psi_N|v_j^X|\Psi_M\rangle$, and $X\in\mathcal X$ as above. The intrinsic OGSR is $\widetilde C_{aj}^{\rm OG}\equiv\widetilde C_{aj}^{\orb,\perp}$; the other three blocks complete the exact response and may themselves contain quantum geometry of the exact Hamiltonian. For momentum-independent SOC the same two-block shorthand applies. The labels $(N,M)$ denote exact-band coherence, whereas $(\orb,\parallel/\perp)$ refer to parent orbital manifolds. The expressions for $C^{\rm OG}$ and $\widetilde C^{\rm OG}$ represent the central results of this work. 

For the purposes of this work I use \textit{OAM-to-spin conversion} in a deliberately narrow modern-theory sense: a spin contribution is classified as conversion when a complete OAM-response kernel reappears with the same occupations and energy denominators, multiplied by an SOC coefficient. The equilibrium moment is $L_{0n}^a=\epsilon_{abc}\sum_{m\ne n}\operatorname{Re}[{\cal R}_{nm}^bv_{mn}^c]$, while~\cite{CullenOME2026} $L_E=L_{\rm OE}+L_{E1}+L_{E2}+L_{E3}$. Here $L_{\rm OE}$ displaces $L_0$ with the Fermi surface, $L_{E1}$ displaces the wave packet, $L_{E2}$ contains connection--derivative and itinerant-circulation structures, and $L_{E3}$ contains three-manifold paths, which vanish in a two-band parent. The seminal atom-centred continuity framework developed in a number of recent papers tracks exchange among electronic spin, local OAM, lattice, and magnetization in heterostructures~\cite{Go2020,Go2021,Jo2024}. Its explicit local OAM operator makes a sequential OAM-to-spin picture natural. The complementary question here is whether the complete modern-theory bulk OAM---including itinerant and local-circulation pieces---reappears inside the spin response with the same weights. The result therefore preserves atom-centred transfer theory while identifying additional orbital-geometric channels and the care required for itinerant Bloch OAM.

The distinction between OGSR and OAM may be seen clearly at weak SOC. We start with the band-diagonal Fermi surface channel. Let $|\chi_{n\zeta}\rangle$ be a spin eigenstate of the band-diagonal SOC $(H_{\so})_{nn}$, with helicity $\zeta$. Standard nondegenerate perturbation theory gives
\begin{equation}
\delta v_{j,n\zeta}^{\orb,\perp}
=\frac{2}{\hbar}\operatorname{Re}\sum_{m\ne n}
i{\cal R}_{nm}^{j}
\langle\chi_{n\zeta}|(H_{\so})_{mn}|\chi_{n\zeta}\rangle.
\label{eq:weak_og_velocity}
\end{equation}
Equation~\eqref{eq:weak_og_velocity} is the covariant-derivative contribution generated by the moving parent-band frame. Writing $h_n=(H_{\so})_{nn}$ and $D_jH_{\so}=\partial_jH_{\so}-i[\mathcal R^j,H_{\so}]$, its right-hand side is $\hbar^{-1}[\partial_jh_n-(D_jH_{\so})_{nn}]$; for momentum-independent SOC it supplies the entire derivative of the projected spin splitting. Hence the leading Fermi-surface OGSR is $O(H_{\so}\tau)$, with no spin-trace cancellation. Its tensor ${\cal R}_{nm}^j(H_{\so})_{mn}$ reproduces the antisymmetric ${\cal R}_{nm}^bv_{mn}^c$ defining $L_0$ only if the microscopic SOC vertex supplies the missing velocity structure pointwise. The response expansion alone therefore imposes no universal quadratic selection rule, although it does not establish first-order OAM conversion for an arbitrary physical SOC~\cite{Supplement}.

For the regular intrinsic channel define $A_{nm}=(H_{\so})_{nm}/(\varepsilon_m-\varepsilon_n)$. To weak SOC, $\widetilde s_a=s_a+[s_a,A]+\cdots$ and $\widetilde{\cal R}_j={\cal R}_j+iD_jA+\cdots$. For an unpolarised spin-degenerate parent pair with equal occupations and a cross-gap denominator finite as $H_{\so}\to0$, the spin trace of every linear term vanishes. The cross-gap OGSR therefore begins with $[s_a,A](iD_jA)=O(H_{\so}^2)$, and any identification with $L_{E1}$, $L_{E2}$, or $L_{E3}$ inherits this onset. The conclusion need not hold for a spin-polarised parent, unequal spin occupations, or nearly degenerate helicities, which form a separate dephasing-sensitive sector~\cite{Supplement}. Under these conditions the inter-band response starts at quadratic order in the SOC, whereas the Fermi-surface OGSR is linear in the SOC and its identification with OAM depends on the SOC vertex.

\textit{Model applications}. I will focus on a generic model of massive Dirac fermions, followed by a concrete model suitable for transition metal dichalcogenides (TMDCs). Peculiarities associated with envelope function Hamiltonians are discussed in the Supplement. Consider a two-band parent $H_p=\epsilon_0+\bd\cdot\bsigma$ and weak $H_{\so}=\sum_c(a_c+\bm b_c\cdot\bsigma)s_c$, where $\bsigma$ is orbital pseudospin, $d=|\bd|$, and $\hat{\bd}=\bd/d$. Projection into band $\eta=\pm$ gives the spin field $\beta_{\eta c}=a_c+\eta\hat{\bd}\cdot\bm b_c$, with $\hat\beta_{\eta c}=\beta_{\eta c}/|\bm\beta_\eta|$. With $\bm b_{c\perp}=\bm b_c-(\hat{\bd}\cdot\bm b_c)\hat{\bd}$, the leading inter-band velocity is $\delta v_{j,\eta\zeta}^{\orb,\perp}=\zeta\hat\beta_{\eta c}{\cal M}_{\eta jc}$, where ${\cal M}_{\eta jc}=(\eta\hbar)^{-1}\partial_j\hat{\bd}\cdot\bm b_{c\perp}$. The two helicity branches $\zeta=\pm$ have opposite spins and opposite velocity corrections, so they contribute with the same sign. The OGSR therefore measures how the SOC vertex couples to changes of the parent orbital texture. If $H_{\so}=\hbar\Lambda_{ic}v_i^{\orb}s_c$, then ${\cal M}_{\eta jc}=2\Delta_\eta\Lambda_{ic}g_{ji}/\hbar$, with $\Delta_\eta=\varepsilon_\eta-\varepsilon_{-\eta}=2\eta d$ and $g_{ji}=\partial_j\hat{\bd}\cdot\partial_i\hat{\bd}/4$. By contrast, the Berry curvature and orbital moment are $\Omega_{\eta,ij}=-(\eta/2)\hat{\bd}\cdot(\partial_i\hat{\bd}\times\partial_j\hat{\bd})$ and $L_\eta^z=-(\Delta_\eta/\hbar)\Omega_{\eta,xy}$. Thus the OGSR samples the symmetric change of the orbital texture, whereas OAM selects its antisymmetric circulation. Although the equilibrium OAM reverses between valleys, the field-induced occupation shift reverses with it; hence both the orbital Edelstein response and the longitudinal Fermi-surface spin response add between valleys. The intrinsic inter-band spin response is instead valley odd and cancels between equally occupied time-reversed valleys. The general weak-SOC tensor is derived in the Supplement.

For a rotationally symmetric conduction band with energy $\varepsilon(k)$ and projected spin splitting $r(k)$, the two Fermi contours give $C_\parallel=-(r+kr')/(2\pi\hbar\varepsilon')$, $C_\perp=kr'/(2\pi\hbar\varepsilon')$, and $C=-r/(2\pi\hbar\varepsilon')$, evaluated at $k_F$ for one valley. The OGSR is the variation of the projected spin splitting across the displaced Fermi surface. On an electron-like contour it opposes $C_\parallel$ whenever $r'(k_F)>0$; the sign is therefore controlled by the evolution of the orbital texture, not by geometry alone. The cancellation also has a transparent origin: $C_\parallel$ contains both the splitting $r$ and its radial variation $kr'$, whereas $C_\perp$ removes precisely the second piece. Only the total $C$ is determined by the splitting itself. Counterflow is therefore generic over a finite region of parameter space, although neither its sign nor an exact cancellation is universal.

I now consider massive Dirac fermions,
\begin{align}
H_\nu={}&Dk^2+\hbar v_{t,\nu}k_x+\alpha(\nu k_x\sigma_x+k_y\sigma_y)
\nonumber\\
&+(m+Bk^2)\sigma_z+\lambda_R(\nu\sigma_xs_y-\sigma_ys_x),
\label{eq:dirac}
\end{align}
where $\nu=\pm$ labels time-reversed valleys, $v_{t,\nu}=\nu v_t$, and $D$ and $B$ bend the scalar dispersion and mass texture. The massless Dirac–Rashba model has also served as an analytic benchmark for multiband quantum kinetic theory~\cite{RaimondiValet2025Dirac}. For $D=B=v_t=0$, the parent conduction-band moment is $L_{+\nu}^z=\nu m\alpha^2/[\hbar(\alpha^2k^2+m^2)]$, but its $E_x$-induced occupation shift integrates to zero. Nevertheless, at weak Rashba SOC, $C_\parallel=-[\lambda_R/(2\pi\hbar\alpha)](1+m^2/\mu^2)$ and $C_\perp=[\lambda_R/(2\pi\hbar\alpha)]m^2/\mu^2$, giving $C=-\lambda_R/(2\pi\hbar\alpha)$. Thus the $O(\lambda_R\tau)$ OGSR cancels the mass-dependent direct response despite zero net OAM, with important experimental implications: doping or gap dependence expected from the direct projected-band spin response can disappear because of orbital-geometric counterflow. The linear tensor is metric-like; the antisymmetric OAM loop first appears through $PH_RQH_RP/(\varepsilon_P-\varepsilon_Q)=C_PP-(\hbar\lambda_R^2/\alpha^2)L_P^zs_z$. Hence identifiable OAM-derived terms begin at $O(\lambda_R^2)$, produce $s_z$, and do not cause the in-plane counterflow.

\begin{figure}[tbp]
\centering
\includegraphics[width=\columnwidth]{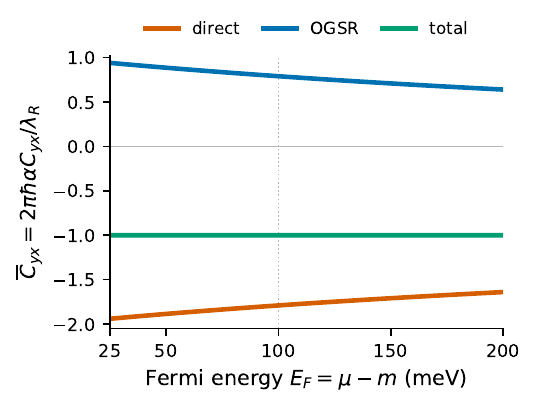}
\caption{Exact $E_x\to s_y$ response for the minimal two-band Dirac–Rashba model parameterized by WSe$_2$ with a linear cone ($D=B=v_t=0$): $m=0.800$ eV, $\alpha=3.939$ eV\,\AA, and $\lambda_R=0.366$ meV at $E_z=10^{-2}$ V/\AA. Here $E_F=\mu-m$, $\overline{C}=2\pi\hbar\alpha C/\lambda_R$. Dotted line $\equiv$ $100$ meV table entry.}
\label{fig:counterflow}
\end{figure}

Figure~\ref{fig:counterflow} shows counterflow throughout the relevant doping window; the Supplement gives the exact expressions and the $D,B\ne0$ result. Table~\ref{tab:tmdc} uses standard TMDC parameters~\cite{Xiao2012,Kormanyos2014}, $E_z=10^{-2}$ V/\AA, and $\lambda_R=(m/\alpha)|\lambda_{\rm BR}|$. At $E_F=100$ meV, the OGSR cancels $93$--$94\%$ of the direct term ($87$--$99\%$ over $25$--$200$ meV)~\cite{Supplement}.

For $D=B=0$, tilt enables $E_x\rightarrow s_z$, and the exact spin texture factorizes through the parent moment,
\begin{equation}
\langle s_z\rangle_{s\nu}=-\frac{s\hbar\lambda_R\varepsilon_k^2}
{\alpha^2{\cal E}_s\sqrt{\alpha^2k^2+\lambda_R^2}}L_{+\nu}^z(k).
\label{eq:oam_factorization}
\end{equation}
Here $\varepsilon_k=(\alpha^2k^2+m^2)^{1/2}$ and ${\cal E}_s=[m^2+(\sqrt{\alpha^2k^2+\lambda_R^2}+s\lambda_R)^2]^{1/2}$. The helicity contributions cancel at linear order, leaving $S_{z,\nu}=e\tau E_xC_L=O(\lambda_R^2\tau)$ with $C_L=\nu v_{t,\nu}m\lambda_R^2/(2\pi\alpha^2\mu^2)+O(v_t\lambda_R^4)$. It adds between valleys and, for $m,\mu>0$, follows the orbital Edelstein response with no competing non-OAM $s_z$ term~\cite{Supplement}. When the Fermi energy is in the conduction band, the inter-band contribution of Eq.~\eqref{eq:intrinsic_general} yields
\begin{equation}
S_{z,{\rm cv},\nu}^{\rm int,(2)}
=-\frac{\hbar\lambda_R^2}{\alpha^2\mu}
(4L_{E1,\nu}^z+2L_{E2,\nu}^z),
\label{eq:intrinsic_conversion}
\end{equation}
with $L_{E3}=0$ in the two-band parent. This is genuine conversion because the occupations and energy weights match. In the gap, however, $L_{E1}=L_{E2}=eE_yv_{t,\nu}/(12\pi m)$ remains finite in each valley while the spin response vanishes: field-induced OAM need not generate spin. Exact-band and dephasing results are given in the Supplement~\cite{Supplement}.

\textit{Discussion}. The central conclusion is that electrical spin and OAM generation are more appropriately viewed as electric field $\rightarrow$ inter-band quantum geometry $\rightarrow\{L_E,S_E\}$, rather than a sequential process of the form electric field $\rightarrow L_E\rightarrow S_E$. The OGSR generically contains an intra-band Fermi-surface contribution and an inter-band Fermi-sea contribution. The former, which is expected to dominate in high-mobility systems, arises essentially from the covariant derivative of the spin-orbit Hamiltonian in the eigenstate basis of the parent Hamiltonian. The exact quantum geometric structure of either contribution depends on the explicit form(s) of SOC present in a device. Consequently, there is no a priori relationship between any component of the OGSR and either the equilibrium or non-equilibrium OAM: spin and OAM densities may descend from the same parent-band coherence without being proportional. For an untilted massive Dirac cone, the OAM generated by the Fermi-surface displacement vanishes after momentum integration while the $O(\lambda_R\tau)$ OGSR remains finite; conversely, the insulating tilted cone has finite valley-resolved $L_{E1}$ and $L_{E2}$ but no intrinsic spin response. Moreover, whereas the OGSR is present at first order in SOC, OAM-related terms may start at higher orders and need not dominate spin generation. For the Dirac cone all identifiable OAM-related terms start at second order in SOC. A large orbital response does not imply a large spin response, and orbital geometry is not by itself evidence of OAM conversion.

\begin{table}[tbp]
\caption{Two-valley coefficients at $E_F=100$ meV and $E_z=10^{-2}$ V/\AA, with $10^4\hbar C$ in nm$^{-1}$, and ${\cal B}=C_\perp/|C_\parallel|$.}
\label{tab:tmdc}
\centering
\small
\setlength{\tabcolsep}{2.2pt}
\begin{tabular}{lrrrrr}
\toprule
 & $\lambda_R$ (meV) & $10^4\hbar C_\parallel$ & $10^4\hbar C_\perp$ & $10^4\hbar C$ & ${\cal B}$\\
\midrule
MoS$_2$  & $0.078$ & $-1.321$ & $1.242$ & $-0.079$ & $0.940$\\
MoSe$_2$ & $0.130$ & $-2.482$ & $2.314$ & $-0.168$ & $0.932$\\
WS$_2$   & $0.266$ & $-3.871$ & $3.654$ & $-0.218$ & $0.944$\\
WSe$_2$  & $0.366$ & $-5.968$ & $5.591$ & $-0.377$ & $0.937$\\
\bottomrule
\end{tabular}
\end{table}

The second conclusion is that SOC does not simply convert orbital dynamics into spin. Orbitronics relies on both strong orbital dynamics and SOC, yet SOC generates spin through several mechanisms whose relative magnitudes must be determined separately. In particular, the familiar direct mechanism, in which SOC shifts the band energies, is present even when the spinless parent is a single, orbitally trivial band. It underlies the current-induced spin polarisation in semiconductor bands described at leading order by a parabolic effective-mass $\bm{k}\cdot\bm{p}$ Hamiltonian with SOC~\cite{Aronov1989,Edelstein1990}. In a more complex system where quantum geometry is important, the spin density resulting from orbital dynamics, with or without OAM, may enhance or suppress the total spin density. The relationship must be determined for each system, particularly because several intrinsic and extrinsic forms of SOC are usually present.

The massive Dirac model makes this separation quantitative: for representative transition-metal dichalcogenide parameters, the leading OGSR cancels $93$--$94\%$ of the direct response. All identifiable OAM-derived spin responses begin at $O(\lambda_R^2)$: the Fermi-surface conversion is $O(\lambda_R^2\tau)$ and the regular intrinsic conversion is $O(\lambda_R^2)$. The dominant $O(\lambda_R\tau)$ OGSR is not OAM and opposes the direct response. The quadratic Fermi-surface OAM term instead produces the orthogonal $s_z$ response and does not cause this counterflow. Thus the leading counterflow is an orbital-geometric correction to spin transport rather than OAM conversion. The quadratic onset is general for the regular inter-band channel but, as Eq.~\eqref{eq:weak_og_velocity} shows, is model dependent for the Fermi-surface channel.


For orbitronic devices, this separation implies that increasing the orbital response alone is not a reliable optimisation strategy. The useful target is the total spin density or torque after the direct, SOC-velocity, OGSR, and genuine OAM-conversion channels have been resolved: the OGSR may reinforce the direct response, but in the Dirac example it produces strong counterflow. Band and interface engineering should therefore identify orbital textures and SOC vertices that suppress counterflow while enhancing the specific OAM tensor that couples to spin.

In summary, I have introduced the OGSR as the part of the electrically induced spin density that probes inter-band coherence of a spinless parent Hamiltonian, and distinguished it from genuine OAM-to-spin conversion by matching the complete OAM tensor, occupations, and energy weights. The massive Dirac model shows that the leading OGSR can survive without net OAM and oppose the direct response, whereas identifiable OAM conversion begins at higher order in SOC. The theory provides a practical framework for determining which orbital channels generate a strong spin density.

\acknowledgments
This work is supported by the Australian Research Council Discovery Project DP2401062.

\renewcommand{\bibfont}{\footnotesize}
\bibliographystyle{unsrt}
\bibliography{references}

\clearpage
\setcounter{equation}{0}
\setcounter{section}{0}
\setcounter{figure}{0}
\setcounter{table}{0}
\renewcommand{\theequation}{S\arabic{equation}}
\renewcommand{\theHequation}{S.\arabic{equation}}
\renewcommand{\thesection}{S\arabic{section}}
\renewcommand{\thefigure}{S\arabic{figure}}
\renewcommand{\thetable}{S\arabic{table}}
\renewcommand{\theHfigure}{S\arabic{figure}}
\renewcommand{\theHtable}{S\arabic{table}}

\begin{widetext}
\begin{center}
{\large\bfseries Supplemental Material for ``Orbital-geometric spin response: inter-band coherence versus orbital-to-spin conversion''\par}
\vspace{1ex}
Dimitrie Culcer\\[2pt]
\small School of Physics, The University of New South Wales, Sydney 2052, Australia
\end{center}
\end{widetext}

\section{Envelope-function and downfolded Hamiltonians}

There is no formal obstruction to applying the OGSR construction to an envelope-function Hamiltonian. The qualification is that downfolding acts on physical operators and perturbations as well as on the Hamiltonian. If $|u_{n\bk}\rangle=\sum_a|\phi_a(\bk)\rangle c_{an}(\bk)$, the retained-subspace connection is
\begin{align}
{\cal R}_{nm}^i
&=ic_n^\dagger\partial_i c_m
+c_n^\dagger {\cal A}^{\rm emb}_i c_m,
\nonumber\\
({\cal A}^{\rm emb}_i)_{ab}
&=i\langle\phi_a|\partial_i\phi_b\rangle.
\label{eq:SM_embedding_connection}
\end{align}
The projected position is $r_i^{\rm eff}=i\partial_i+{\cal A}^{\rm emb}_i$, and therefore $v_i^{\rm eff}=\hbar^{-1}(\partial_iH_{\rm eff}-i[{\cal A}^{\rm emb}_i,H_{\rm eff}])$. Hence $\hbar^{-1}\partial_iH_{\rm eff}$ is the physical velocity only in a fixed momentum-independent basis, or when the embedding term has already been included. More generally every operator and driving term must be transformed as $O_{\rm eff}=PU^\dagger OUP$; applying the downfolding transformation only to the Hamiltonian can change spin--orbit response functions~\cite{AdoPosition2024,AdoKubo2023}.

The massive-Dirac calculation defines such a fixed-basis effective Bloch model, so its counterflow result is exact within the model. Its OAM is nevertheless the active-subspace OAM: remote bands may renormalise the moment and restore $L_{E3}$. Matching a projected conduction-band Rashba coefficient also fixes the band-edge splitting but not generally $P_+H_{\so}P_-$, which controls the OGSR. Quantitative material predictions therefore require a multiband $\bm{k}\cdot\bm{p}$ or Wannier model with position, velocity, spin, OAM, and SOC vertices projected consistently.

The calculation assumes an infinite homogeneous system. In an inhomogeneous envelope description, gradients of the band-edge basis and material parameters generate interface terms~\cite{Foreman1996}. Boundary responses absent from the bulk include reflection-induced orbital magnetisation, spin-filtering magnetoresistance, and Kerr polarisation~\cite{Voss2025,IorshTitov2026,IorshTitovKerr2026}. An edge density cannot be inferred from the bulk Hamiltonian. Boundary OGSR requires consistent projected operators, interface terms, and a corresponding real-space extension.

\section{Projector generalisation for composite bands}

For isolated nondegenerate bands the response reduces to the scalar matrix elements in the Letter. Projectors group degenerate or entangled bands without selecting an internal basis. The parent is fixed by adiabatically removing microscopic SOC while retaining the orbital basis and spin-independent terms; it is not an arbitrary repartition of $H$.

Let
\begin{align}
H_{\orb}&=\sum_A\varepsilon_AP_A,
\\
H=H_{\orb}+H_{\so}&=\sum_NE_N\Pi_N .
\label{eq:SM_spectral}
\end{align}
The two sets of spectral projectors obey
\begin{align}
P_AP_B&=\delta_{AB}P_A,& \sum_AP_A&=1,
\\
\Pi_N\Pi_M&=\delta_{NM}\Pi_N,& \sum_N\Pi_N&=1.
\end{align}
The $P_A$ define parent orbital manifolds and include the identity in real spin. The $\Pi_N$ define the exact spin--orbit-coupled manifolds. Both can have rank greater than one. Bands connected by a crossing after removing SOC belong to one composite $P_A$; subdivision at the crossing is not invariant. For distinct parent and exact manifolds define the operator-valued Berry connections
\begin{align}
{\cal R}_{AB}^{0,i}&=iP_A\partial_iP_B,
& A&\ne B,
\\
{\cal R}_{NM}^{i}&=i\Pi_N\partial_i\Pi_M,
& N&\ne M.
\label{eq:SM_connections}
\end{align}
For rank-one manifolds their coefficients reduce to the usual ${\cal R}_{mn}^i=i\langle u_m|\partial_i u_n\rangle$. The superscript $0$ distinguishes the parent connection from the exact connection throughout.
For an arbitrary $H_{\so}(\bk)$ define
\begin{align}
v_i^{\chi,\parallel}&=\sum_AP_Av_i^\chi P_A,
&
v_i^{\chi,\perp}&=\sum_{A\neq B}P_Av_i^\chi P_B,
\\
v_i^\chi&=\hbar^{-1}\partial_iH_\chi,
&
\chi&\in\{\orb,\so\}.
\label{eq:SM_four_blocks}
\end{align}
Hence
\begin{equation}
v_i=v_i^{\orb,\parallel}+v_i^{\orb,\perp}
+v_i^{\so,\parallel}+v_i^{\so,\perp}.
\label{eq:SM_total_four_blocks}
\end{equation}
The two spin--orbit blocks in Eq.~\eqref{eq:SM_four_blocks} are retained in the exact response. They are not included in the OGSR because the identity derived next follows from the eigenvalue equation for $H_{\orb}$ and has no analogue for a general $H_{\so}$.

Differentiating $H_{\orb}P_B=\varepsilon_BP_B$ gives
\begin{equation}
(\partial_iH_{\orb})P_B+H_{\orb}(\partial_iP_B)
=(\partial_i\varepsilon_B)P_B+\varepsilon_B(\partial_iP_B).
\end{equation}
Multiplication by $P_A$ from the left, with $A\neq B$, yields
\begin{equation}
P_A(\partial_iH_{\orb})P_B
=(\varepsilon_B-\varepsilon_A)P_A(\partial_iP_B),
\end{equation}
and hence
\begin{equation}
P_Av_i^{\orb}P_B
=\frac{\varepsilon_B-\varepsilon_A}{\hbar}P_A(\partial_iP_B).
\label{eq:SM_velocity_identity}
\end{equation}
Equivalently,
\begin{equation}
v_i^{\orb}
=\frac{1}{\hbar}\sum_A(\partial_i\varepsilon_A)P_A
+\frac{i}{\hbar}\sum_{A\ne B}
(\varepsilon_A-\varepsilon_B){\cal R}_{AB}^{0,i}.
\label{eq:SM_velocity_connection}
\end{equation}
The diagonal and off-diagonal terms in Eq.~\eqref{eq:SM_velocity_connection} are separately Hermitian. Indeed,
\begin{equation}
({\cal R}_{AB}^{0,i})^\dagger={\cal R}_{BA}^{0,i},
\end{equation}
so the term for $(B,A)$ is the Hermitian conjugate of the term for $(A,B)$. For exact manifolds the same reasoning gives
\begin{equation}
\Pi_Nv_i\Pi_M
=\frac{i}{\hbar}(E_N-E_M){\cal R}_{NM}^{i},
\quad N\ne M.
\label{eq:SM_exact_velocity_connection}
\end{equation}

With the ordered-pair convention $v_i^{\orb,\perp}=\sum_{A\neq B}P_Av_i^{\orb}P_B$, the precise identity is
\begin{equation}
v_i^{\orb,\perp}
=\frac{1}{2}\sum_A\left[P_A,[P_A,v_i^{\orb}]\right].
\label{eq:SM_double_commutator_precise}
\end{equation}
Indeed, the term associated with each ordered pair $(A,B)$ occurs once from $A$ and once from $B$ in the double-commutator sum.

Equations~\eqref{eq:SM_velocity_identity} and \eqref{eq:SM_double_commutator_precise} contain only the spectral projectors. A change of basis inside a composite parent manifold,
\begin{equation}
|u_{Aa}\rangle\rightarrow\sum_b|u_{Ab}\rangle U^A_{ba}(\bk),
\end{equation}
does not change
\begin{equation}
P_A=\sum_a|u_{Aa}\rangle\langle u_{Aa}|.
\end{equation}
It therefore leaves $v_i^{\orb,\parallel}$, $v_i^{\orb,\perp}$, and every response defined from them unchanged. This is the relevant non-Abelian gauge covariance for a fixed parent and grouping, not invariance under repartitioning $H$ or regrouping across a finite parent gap.

In practice, scalar- and fully relativistic Hamiltonians are constructed in the same Wannier gauge, with $H_{\orb}=H_{\rm scalar}$ and $H_{\so}=H_{\rm full}-H_{\rm scalar}$; the selected groups define $P_A(\bk)$.

For an exact spectral projector $\Pi_N$ define
\begin{equation}
\Gamma^N_{BA}=P_B\Pi_NP_A.
\end{equation}
Using cyclicity of the trace and Eq.~\eqref{eq:SM_velocity_identity},
\begin{align}
\tr(\Pi_Nv_i^{\orb,\perp})
&=\sum_{A\neq B}\Tr(\Pi_NP_Av_iP_B)
\nonumber\\
&=\frac{1}{\hbar}\sum_{A\neq B}
(\varepsilon_B-\varepsilon_A)
\Tr\left[\Gamma^N_{BA}(\partial_iP_B)\right].
\label{eq:SM_exact_geometry}
\end{align}
The Hermitian-conjugate ordered pair makes the sum explicitly real.

For the leading band-diagonal Fermi-surface response in the RTA, the gauge-covariant trace formula is
\begin{equation}
(\rho_E^{d})_{NN}=e\tau E_i v_{iN}(-f_N'),
\quad
v_{iN}=\langle\Psi_N|v_i|\Psi_N\rangle,
\label{eq:SM_diagonal_dm}
\end{equation}
where $f_N=f(E_N)$, $f_N'=\partial f(E_N)/\partial E_N$, and
$\int_{\bk}=\int\dd^dk/(2\pi)^d$.  For isolated nondegenerate exact
bands this gives
\begin{align}
S_a^{\rm FS}&=e\tau E_i
\big(C_{ai}^{\parallel}+C_{ai}^{\perp}\big),
\nonumber\\
C_{ai}^{X}&=\sum_N\int_{\bk}(-f_N')s_{aN}v_{iN}^{X},
\quad X=\parallel,\perp,
\label{eq:SM_nondeg_response}
\end{align}
with $s_{aN}=\langle\Psi_N|s_a|\Psi_N\rangle$ and
$v_{iN}^{X}=\langle\Psi_N|v_i^X|\Psi_N\rangle$.  Equation~\eqref{eq:SM_nondeg_response}
is the rank-one form of
\begin{equation}
C_{ai}^{X}=\sum_N\int_{\bk}(-f'_N)
\Tr\!\left[s_a\Pi_Nv_i^X\Pi_N\right].
\label{eq:SM_response_trace}
\end{equation}
The four values of $X$ are the four blocks in Eq.~\eqref{eq:SM_total_four_blocks}. The OGSR is the contribution $X=(\orb,\perp)$. Define
\begin{equation}
\Gamma^{N,a}_{BA}=P_B\Pi_Ns_a\Pi_NP_A.
\end{equation}
Substitution of Eq.~\eqref{eq:SM_velocity_identity} into Eq.~\eqref{eq:SM_response_trace} yields
\begin{align}
C_{ai}^{\rm OG}
=\frac{1}{\hbar}\sum_N\int_{\bk}(-f'_N)
\sum_{A\neq B}(\varepsilon_B-\varepsilon_A)
\Tr\!\left[\Gamma^{N,a}_{BA}\partial_iP_B\right].
\label{eq:SM_exact_spin_geometry}
\end{align}
Equation~\eqref{eq:SM_exact_spin_geometry} is valid for rank-one or composite exact manifolds. In the rank-one case, $\Pi_Ns_a\Pi_N=s_{aN}\Pi_N$, so $\Gamma^{N,a}_{BA}=s_{aN}\Gamma^N_{BA}$. The expression then becomes the exact spin expectation multiplied by Eq.~\eqref{eq:SM_exact_geometry}. This explicitly shows why a product of two separately traced expectation values should not be used for a degenerate exact manifold.
The corresponding Pauli-spin polarisation is $S_a^{\rm OG}=e\tau E_iC_{ai}^{\rm OG}$. It is counterflow only when it opposes $C_{ai}^{(\orb,\parallel)}$; projector algebra fixes neither sign nor magnitude. Only the sum is observable. The RTA omits scattering-induced interband coherence and vertex corrections, which can alter the pieces.

\section{Exact clean band-intrinsic response and parent resolution}

The clean band-intrinsic off-diagonal driving term of Culcer, Sekine, and MacDonald [Phys. Rev. B \textbf{96}, 035106 (2017)] is
\begin{equation}
(D_E)_{NM}=\frac{ieE_i}{\hbar}{\cal R}_{NM}^i(f_N-f_M),
\quad N\ne M.
\end{equation}
Here ``intrinsic'' denotes this clean band-coherence contribution. Disorder-generated terms of order $\tau^0$, including scattering-induced interband coherence, are outside the present decomposition.
The principal-value solution of the steady-state Liouville equation is
\begin{equation}
(S_E^{\rm int})_{NM}
=eE_i\frac{f_N-f_M}{E_N-E_M}{\cal R}_{NM}^i.
\label{eq:SM_SE_intrinsic}
\end{equation}
It is Hermitian because ${\cal R}_{MN}=({\cal R}_{NM})^\dagger$ and the divided difference is symmetric under $N\leftrightarrow M$.

For a nearly degenerate pair the clean principal-value solution must not be continued to zero splitting.  A useful diagnostic is obtained by retaining a phenomenological interband dephasing rate $\gamma_{NM}$ in the off-diagonal kinetic equation,
\begin{align}
\left(\frac{i\Delta_{NM}}{\hbar}+\gamma_{NM}\right)
(S_E)_{NM}&=(D_E)_{NM},
\\
\Delta_{NM}&=E_N-E_M .
\end{align}
Writing $\Gamma_{NM}=\hbar\gamma_{NM}$ gives
\begin{equation}
(S_E)_{NM}=eE_i(f_N-f_M){\cal R}_{NM}^i
\frac{\Delta_{NM}+i\Gamma_{NM}}
{\Delta_{NM}^2+\Gamma_{NM}^2}.
\label{eq:SM_SE_dephasing}
\end{equation}
For an ordered band pair its contribution to the spin density is therefore
\begin{align}
S_{a,NM}^{(\gamma)}={}&
2e\hbar E_i(f_N-f_M)
\frac{1}{\Delta_{NM}^2+\Gamma_{NM}^2}
\Bigg\{\operatorname{Im}{\cal V}_{NM}^{ai}
\nonumber\\[-2pt]
&+\frac{\Gamma_{NM}}{\Delta_{NM}}
\operatorname{Re}{\cal V}_{NM}^{ai}\Bigg\},
\label{eq:SM_spin_dephasing}\\[-2pt]
{\cal V}_{NM}^{ai}={}&\Tr[\Pi_Ms_a\Pi_Nv_i\Pi_M].
\end{align}
The first term is the dephasing-broadened continuation of the clean mixed-curvature response: its clean denominator $\Delta_{NM}^{-2}$ is replaced by $(\Delta_{NM}^2+\Gamma_{NM}^2)^{-1}$.  The second is a dissipative term generated by the same scalar-dephasing ansatz.  Equations~\eqref{eq:SM_SE_dephasing} and \eqref{eq:SM_spin_dephasing} are not a substitute for a microscopic disorder collision operator; they only display the noncommuting clean and zero-splitting limits.  A full disorder treatment also contains scattering-induced interband coherence.

Tracing the clean solution with $s_a$ gives
\begin{align}
S_a^{\rm int}=eE_i\sum_{N\ne M}\int_{\bk}
\frac{f_N-f_M}{E_N-E_M}
\Tr[\Pi_Ms_a{\cal R}_{NM}^i].
\label{eq:SM_Sint_R}
\end{align}
Using Eq.~\eqref{eq:SM_exact_velocity_connection}
and combining conjugate ordered pairs gives
\begin{align}
S_a^{\rm int}&=e\hbar E_i\widetilde C_{ai},
\nonumber\\
\widetilde C_{ai}={}&2\sum_{N<M}\int_{\bk}
\frac{f_N-f_M}{(E_N-E_M)^2}
\nonumber\\
&\times\operatorname{Im}\Tr[\Pi_Ms_a\Pi_Nv_i\Pi_M].
\label{eq:SM_Sint_v}
\end{align}

Substitution of the four velocity blocks in Eq.~\eqref{eq:SM_total_four_blocks} gives $\widetilde C_{ai}=\sum_X\widetilde C_{ai}^X$ and $S_a^{\rm int}=\sum_XS_a^{\rm int,X}$, where $\widetilde C_{ai}^X$ is obtained by replacing $v_i$ with $v_i^X$ in Eq.~\eqref{eq:SM_Sint_v}. For the parent-off-diagonal orbital block, Eq.~\eqref{eq:SM_velocity_identity} yields
\begin{align}
S_a^{\rm int,OG}
=2eE_i\sum_{N<M}\int_{\bk}
\frac{f_N-f_M}{(E_N-E_M)^2}
\sum_{A\ne B}(\varepsilon_B-\varepsilon_A)
\nonumber\\[-2pt]
\times\operatorname{Im}\Tr[
\Pi_Ms_a\Pi_NP_A(\partial_iP_B)\Pi_M].
\label{eq:SM_Sint_parent}
\end{align}
This expression is invariant under unitary changes of frame inside any parent or exact composite manifold. In contrast with the Edelstein result, the exact projectors adjacent to $s_a$ are different. The intrinsic response therefore probes field-induced exact-band coherence in addition to the parent coherence contained in each $\Pi_N$.

\section{Weak spin--orbit expansion}

This section is not required to define the exact OGSR. It records how the general projector result reduces to the nondegenerate perturbative language used in the Letter and in earlier density-matrix treatments. For an isolated parent orbital band $n$, let $|\chi_{n\zeta}\rangle$ diagonalize the $2\times2$ spin matrix $(H_{\so})_{nn}$. The first-order correction to the parent-interband velocity is
\begin{align}
\delta v_{i,n\zeta}^{\perp}
&=2\operatorname{Re}\sum_{m\ne n}
\frac{(v_i^{\orb})_{nm}
\langle\chi_{n\zeta}|(H_{\so})_{mn}|\chi_{n\zeta}\rangle}
{\varepsilon_n-\varepsilon_m}
\nonumber\\
&=\frac{2}{\hbar}\operatorname{Re}\sum_{m\ne n}
i{\cal R}_{nm}^{i}
\langle\chi_{n\zeta}|(H_{\so})_{mn}|\chi_{n\zeta}\rangle.
\label{eq:SM_nondeg_vperp}
\end{align}
The first line follows from the first-order correction to the parent-band eigenstate, while the second follows from $(v_i^{\orb})_{nm}=i(\varepsilon_n-\varepsilon_m){\cal R}_{nm}^{i}/\hbar$. Thus the virtual-state energy denominator cancels, leaving the connection term required by the covariant momentum derivative of the SOC matrix in the moving parent-band frame. This is the rank-one form of the operator expression below. It contains one spatial interband connection and one SOC matrix element; no composite-band projectors are needed for the massive-Dirac calculation.

Split the spin-orbit Hamiltonian into parent-block-diagonal and block-off-diagonal parts,
\begin{align}
H_{\so}&=H_{\so}^{\parallel}+H_{\so}^{\perp},
\\
H_{\so}^{\perp}&=\sum_{A\neq B}P_AH_{\so}P_B.
\end{align}
For a state evolving from parent manifold $A$, the leading intermanifold coherence is
\begin{equation}
P_B|\Psi_N\rangle
=\frac{P_BH_{\so}P_A}{\varepsilon_A-\varepsilon_B}
P_A|\Psi_N\rangle+O[(H_{\so}^{\perp})^2].
\label{eq:SM_state_coherence}
\end{equation}
It follows that
\begin{align}
\langle v_i^{\orb,\perp}\rangle_N
&=2\operatorname{Re}\sum_{B\neq A}
\frac{1}{\varepsilon_A-\varepsilon_B}
\nonumber\\
&\quad\times
\langle\psi_A|(v_i^{\orb})_{AB}(H_{\so})_{BA}|\psi_A\rangle
+O[(H_{\so}^{\perp})^2].
\label{eq:SM_vperp_weak_velocity}
\end{align}
Using Eq.~\eqref{eq:SM_velocity_identity}, this becomes
\begin{align}
\langle v_i^{\orb,\perp}\rangle_N
=\frac{i}{\hbar}\sum_{B\neq A}
\big\langle\psi_A\big|
\big[&{\cal R}_{AB}^i(H_{\so})_{BA}
\nonumber\\
&-(H_{\so})_{AB}{\cal R}_{BA}^i\big]
\big|\psi_A\big\rangle,
\label{eq:SM_mixed_tensor}
\end{align}
where ${\cal R}_{AB}^i=iP_A\partial_iP_B$ in a local frame. This quantity combines the parent interband connection with the spin--orbit matrix element. It reduces to the standard quantum-geometric tensor only when that matrix element supplies a second velocity or projector derivative.

For example, if
\begin{equation}
(H_{\so})_{BA}=\sum_{jc}\lambda_{jc}
(v_j^{\orb})_{BA}s_c,
\end{equation}
then Eq.~\eqref{eq:SM_mixed_tensor} contains
\begin{equation}
{\cal R}_{AB}^i{\cal R}_{BA}^j
=g_{ij}^{AB}-\frac{i}{2}\Omega_{ij}^{AB}.
\end{equation}
The symmetric real part gives the quantum metric and the antisymmetric imaginary part gives the Berry curvature. Neither contribution has a fixed sign after contraction with the spin texture and the coefficients $\lambda_{jc}$. Equation~\eqref{eq:SM_mixed_tensor} has no first-order spin-trace cancellation. Thus the Fermi-surface OGSR is $O(H_{\so}\tau)$. Conversion at that order would additionally require the physical SOC vertex to select the complete antisymmetric product with the occupations and weights of $L_0$ or $L_{\rm OE}$. The response expansion supplies no universal quadratic selection rule, but does not by itself prove that a symmetry-allowed first-order conversion occurs for arbitrary SOC.

For the intrinsic response, both vertices must be transformed. Let $e^A$ block diagonalize the Hamiltonian to first order, with
\begin{align}
A_{AB}&=\frac{P_AH_{\so}P_B}{\varepsilon_B-\varepsilon_A},
\\
A_{BA}&=-A_{AB}^\dagger.
\end{align}
The spin and connection vertices in the block-diagonal frame are
\begin{align}
\widetilde s_a
&=s_a+[s_a,A]+\frac12[[s_a,A],A]+O(A^3),
\nonumber\\
\widetilde{\cal R}_i
&={\cal R}_i+iD_iA+\frac{i}{2}[D_iA,A]+O(A^3).
\label{eq:SM_transformed_vertices}
\end{align}
For a regular cross-gap pair the zeroth-order spin vertex is parent-block diagonal. At first order $[s_a,A]_{AB}$ is spin traceless for an unpolarised, equally occupied parent doublet. The first surviving product, such as $[s_a,A](iD_iA)$, is therefore $O(H_{\so}^2)$. This assumes a finite parent gap and excludes spin-polarised or unequally occupied parents and SOC-split helicity pairs.

For velocity-like spin--orbit coupling,
\begin{align}
H_{\so}^{\perp}&=\lambda_{\mu b}s_\mu v_b^{\orb,\perp},
\\
A_{AB}&=\frac{i}{\hbar}\lambda_{\mu b}s_\mu{\cal R}_{AB}^b.
\label{eq:SM_A_velocity}
\end{align}
The product $[s_a,A](iD_iA)$ contains
\begin{align}
\epsilon_{a\mu\nu}\lambda_{\mu b}\lambda_{\nu c}
\operatorname{Re}\Tr[{\cal R}_{AB}^c(D_i{\cal R}^b)_{BA}]
\nonumber\\
=(\operatorname{cof}\lambda)_{a\ell}\epsilon_{\ell bc}
\operatorname{Re}\Tr[{\cal R}_{AB}^c(D_i{\cal R}^b)_{BA}],
\label{eq:SM_T_tensor}
\end{align}
which is the connection--derivative tensor in the first term of $L_{E2}$. The double commutators in Eq.~\eqref{eq:SM_transformed_vertices} generate three-manifold products ${\cal R}{\cal R}v$, while expansion of the exact divided difference and the diagonal blocks of the transformed Hamiltonian generates the velocity-sum products in $L_{E1}$ and in the second term of $L_{E2}$. For a two-parent-band model the genuine three-manifold term vanishes, consistently with $L_{E3}=0$.

For this velocity-like vertex the quadratic products map explicitly onto modern-theory OAM tensors. For general SOC the regular quadratic onset remains, but conversion requires the complete OAM occupations and weights. SOC-resolved helicity coherence is separate from Eq.~\eqref{eq:SM_T_tensor}.

\section{Generic two-band parent and the OGSR}
\label{sec:SM_generic_two_band}

The two-level structure in this section is an orbital or other \emph{pseudospin}; it is neither the physical spin nor the valley degree of freedom.  We therefore use $\bsigma$ for pseudospin, $s_c$ for real spin, and an explicit label $\nu=\pm$ for valleys.  The momentum $\bq$ is measured from the centre of valley $\nu$.  We first assume that the Hamiltonian is valley diagonal, so that $\nu$ is a good quantum number.  The generic spin-independent parent is
\begin{equation}
H_{p,\nu}(\bq)=\epsilon_{0\nu}(\bq)1+\bd_\nu(\bq)\cdot\bsigma,
\quad
P_{\eta\nu}=\frac{1}{2}(1+\eta\hat{\bd}_\nu\cdot\bsigma),
\label{eq:SM_generic_parent_valley}
\end{equation}
with $\eta=\pm$, $d_\nu=|\bd_\nu|$, $\hat{\bd}_\nu=\bd_\nu/d_\nu$, and parent energies $\varepsilon_{\eta\nu}=\epsilon_{0\nu}+\eta d_\nu$.  The most general \emph{valley-conserving} weak Hermitian coupling linear in real spin can be written, with real coefficient functions,
\begin{equation}
H_{\so,\nu}(\bq)=\sum_c\left[a_{c\nu}(\bq)1+\bb_{c\nu}(\bq)\cdot\bsigma\right]s_c.
\label{eq:SM_generic_SOC}
\end{equation}
The choice of $a_{c\nu}$ and $\bb_{c\nu}$ is part of the microscopic spin--orbit model and is not fixed by the pseudospin parent.  Valley-dependent Ising and Rashba couplings are both contained in Eq.~\eqref{eq:SM_generic_SOC}; intervalley spin--orbit coupling is discussed below.

\subsection{Global crystal momentum and local valley coordinates}

Because several valley signs below depend on whether a derivative is taken with respect to the global crystal momentum or the local coordinate, we state the convention explicitly.  Let $\bm K_\nu$ denote the centre of valley $\nu$ and write
\begin{equation}
\bk=\bm K_\nu+\bq.
\label{eq:SM_global_local_k}
\end{equation}
For a time-reversal pair, $\bm K_{-\nu}=-\bm K_\nu$ modulo a reciprocal lattice vector.  Hence the time-reversed partner of the state $(\nu,\bq)$ has global momentum $-\bk$ and local coordinate $-\bq$ in valley $-\nu$:
\begin{equation}
\bm K_\nu+\bq\xrightarrow{\Theta}
-(\bm K_\nu+\bq)\equiv \bm K_{-\nu}-\bq.
\label{eq:SM_global_TR_pair}
\end{equation}
Within a fixed valley, $\partial_{k_i}=\partial_{q_i}$, but when a relation between the two time-reversed valleys is differentiated the argument $-\bq$ produces an additional minus sign.  This elementary point is responsible for the parity of the nonequilibrium Fermi-surface occupation derived explicitly below.  We therefore never identify the pair $(\nu,\bq)$ with $(-\nu,\bq)$; the paired point is always $(-\nu,-\bq)$.

\subsection{Time-reversal bookkeeping}

When $\nu=\pm$ are a time-reversal pair, the valley labels and the local momenta transform together,
\begin{equation}
(\nu,\bq)\xrightarrow{\Theta}(-\nu,-\bq).
\label{eq:SM_TR_valley_map}
\end{equation}
In the pseudospin basis used for the Dirac models below, spinless time reversal acts by complex conjugation while exchanging the valleys.  It is useful to introduce
\begin{equation}
R_T=\operatorname{diag}(1,-1,1),
\quad
K\bsigma K^{-1}=R_T\bsigma,
\end{equation}
where $K$ denotes complex conjugation.  Including real spin, $\Theta=i s_yK$ together with the implicit valley exchange, so that $\Theta s_c\Theta^{-1}=-s_c$.  The condition
\begin{equation}
\Theta H_\nu(\bq)\Theta^{-1}=H_{-\nu}(-\bq)
\label{eq:SM_TR_condition}
\end{equation}
then gives
\begin{align}
\epsilon_{0,-\nu}(-\bq)&=\epsilon_{0\nu}(\bq),
&\bd_{-\nu}(-\bq)&=R_T\bd_\nu(\bq),
\label{eq:SM_TR_parent}\\[2pt]
a_{c,-\nu}(-\bq)&=-a_{c\nu}(\bq),
&\bb_{c,-\nu}(-\bq)&=-R_T\bb_{c\nu}(\bq).
\label{eq:SM_TR_SOC}
\end{align}
These relations are not imposed if the two valleys are not related by time reversal.  Notice in particular that a coefficient multiplying the identity in pseudospin is valley odd when it multiplies real spin, whereas a $\sigma_y s_c$ coefficient is valley even in this basis.  This is the origin of the different explicit valley factors in the Ising and Rashba terms used below.

\subsubsection{Equivalent formulation in an explicit valley space}

The block notation above is sufficient whenever valley is conserved, but it is useful to state its relation to a genuine valley Hilbert space because this removes any possible ambiguity between valley and pseudospin.  Introduce Pauli matrices $\tau_\mu^{\rm v}$, $\mu=0,x,y,z$, acting only on the $(+,-)$ valley doublet.  In a reduced-zone representation compatible with the Dirac basis used here, time reversal may be written
\begin{equation}
\Theta=\tau_x^{\rm v}\otimes 1_\sigma\otimes i s_y K.
\label{eq:SM_TR_full_valley_operator}
\end{equation}
The three factors act respectively in valley, pseudospin, and real-spin space.  With the phase convention of Eq.~\eqref{eq:SM_TR_full_valley_operator},
\begin{align}
\Theta\tau_\mu^{\rm v}\Theta^{-1}
&=\chi_\mu^{\rm v}\tau_\mu^{\rm v},
& (\chi_0^{\rm v},\chi_x^{\rm v},\chi_y^{\rm v},\chi_z^{\rm v})
&=(+,+,+,-),
\\
\Theta\sigma_b\Theta^{-1}
&=\chi_b^\sigma\sigma_b,
& (\chi_0^\sigma,\chi_x^\sigma,\chi_y^\sigma,\chi_z^\sigma)
&=(+,+,-,+),
\\
\Theta s_c\Theta^{-1}&=-s_c.
\label{eq:SM_TR_parity_table}
\end{align}
The apparently unusual even parity of $\tau_y^{\rm v}$ follows from the two operations in time reversal: complex conjugation changes $\tau_y^{\rm v}\to-\tau_y^{\rm v}$ and the subsequent valley exchange by $\tau_x^{\rm v}$ changes the sign once more.

A completely general Hermitian spin-linear coupling in the enlarged space can be expanded as
\begin{equation}
H_{\so}(\bk)
=\sum_{\mu b c}\lambda_{\mu bc}(\bk)
\tau_\mu^{\rm v}\sigma_b s_c,
\label{eq:SM_full_valley_SOC}
\end{equation}
with real coefficients in this Hermitian matrix basis.  Time reversal requires
\begin{equation}
\lambda_{\mu bc}(-\bk)
=-\chi_\mu^{\rm v}\chi_b^\sigma\,
\lambda_{\mu bc}(\bk).
\label{eq:SM_full_valley_SOC_TR}
\end{equation}
The analogous spin-independent parent coefficient $h_{\mu b}$ obeys
$h_{\mu b}(-\bk)=\chi_\mu^{\rm v}\chi_b^\sigma h_{\mu b}(\bk)$.
Valley conservation corresponds to retaining only $\mu=0,z$; these two components are precisely equivalent to allowing independent but time-reversal-constrained coefficients in the $\nu=\pm$ blocks of Eqs.~\eqref{eq:SM_generic_parent_valley} and \eqref{eq:SM_generic_SOC}.  Terms with $\mu=x,y$ are genuine intervalley couplings.  In an unfolded pristine crystal they require a perturbation carrying the large valley momentum transfer (or a superlattice which folds the valleys together), and they cannot be represented by a mere valley label $\nu$.  They are therefore excluded from the two-band reduction below, but not from the exact projector formalism itself.

As useful checks, a momentum-independent Ising term $\tau_z^{\rm v}(\sigma_z-1)s_z$ is time-reversal even, as is the Rashba combination
$\tau_z^{\rm v}\sigma_xs_y-\tau_0^{\rm v}\sigma_ys_x$.
These are the operator forms of the explicit $\nu$ factors used later.

Projection of Eq.~\eqref{eq:SM_generic_SOC} into parent band $(\eta,\nu)$ gives the effective real-spin field
\begin{equation}
P_{\eta\nu}H_{\so,\nu}P_{\eta\nu}
=P_{\eta\nu}\,\bbeta_{\eta\nu}\cdot\bs\,P_{\eta\nu},
\quad
\beta_{\eta\nu c}=a_{c\nu}+\eta\hat{\bd}_\nu\cdot\bb_{c\nu}.
\label{eq:SM_beta_eta}
\end{equation}
Writing $r_{\eta\nu}=|\bbeta_{\eta\nu}|$ and $\hat{\bbeta}_{\eta\nu}=\bbeta_{\eta\nu}/r_{\eta\nu}$, the two weak-spin--orbit branches $\zeta=\pm$ satisfy
\begin{equation}
E_{\eta\nu\zeta}=\varepsilon_{\eta\nu}+\zeta r_{\eta\nu}+O(H_{\so}^2),
\quad
\langle s_c\rangle_{\eta\nu\zeta}=\zeta\hat\beta_{\eta\nu c}+O(H_{\so}).
\label{eq:SM_generic_helicity}
\end{equation}
Equations~\eqref{eq:SM_TR_parent} and \eqref{eq:SM_TR_SOC} imply
\begin{equation}
r_{\eta,-\nu}(-\bq)=r_{\eta\nu}(\bq),
\quad
\hat{\bbeta}_{\eta,-\nu}(-\bq)=-\hat{\bbeta}_{\eta\nu}(\bq),
\label{eq:SM_TR_beta}
\end{equation}
as required for a time-reversed spin texture.

The block-off-diagonal part of Eq.~\eqref{eq:SM_generic_SOC} is controlled only by the component of $\bb_{c\nu}$ transverse to the parent pseudospin,
\begin{equation}
\bb_{c\nu\perp}=\bb_{c\nu}
-(\hat{\bd}_\nu\cdot\bb_{c\nu})\hat{\bd}_\nu.
\end{equation}
Using first-order interband coherence, or equivalently Eq.~\eqref{eq:SM_mixed_tensor}, the parent-off-diagonal \emph{orbital} velocity in branch $(\eta,\nu,\zeta)$ is
\begin{align}
\delta v_{i,\eta\nu\zeta}^{\orb,\perp}
&=2\operatorname{Re}
\frac{\langle\eta\nu|v_{i,\nu}^{\orb}|\bar\eta\nu\rangle
\langle\bar\eta\nu|H_{\so,\nu}|\eta\nu\rangle_\zeta}
{\varepsilon_{\eta\nu}-\varepsilon_{\bar\eta\nu}}
\nonumber\\
&=\zeta\hat\beta_{\eta\nu c}\,{\cal M}_{\eta\nu ic},
\label{eq:SM_generic_vperp}\\
{\cal M}_{\eta\nu ic}
&=\frac{1}{\eta\hbar}
\partial_{q_i}\hat{\bd}_\nu\cdot\bb_{c\nu\perp}.
\label{eq:SM_generic_mixed_tensor}
\end{align}
Repeated real-spin indices are summed.  At the time-reversed point,
\begin{align}
[\partial_{q_i}\hat{\bd}_{-\nu}]_{-\bq}
&=-R_T[\partial_{q_i}\hat{\bd}_{\nu}]_{\bq},
\\
\bb_{c,-\nu\perp}(-\bq)
&=-R_T\bb_{c\nu\perp}(\bq).
\end{align}
so that
\begin{equation}
{\cal M}_{\eta,-\nu,ic}(-\bq)
={\cal M}_{\eta\nu ic}(\bq).
\label{eq:SM_TR_M}
\end{equation}
Thus the mixed tensor itself is valley even, while the branch spin texture and the corresponding velocity correction are valley odd.  Their product is valley even.  To leading order, when both helicity branches cross the Fermi level,
\begin{equation}
C_{ai,\nu}^{\rm Ed,\perp}
=2\sum_\eta\int_{\bq}(-f'_{\eta\nu})
\hat\beta_{\eta\nu a}\hat\beta_{\eta\nu c}
{\cal M}_{\eta\nu ic}
+\text{higher orders},
\label{eq:SM_generic_Cperp}
\end{equation}
and for equally occupied time-reversal partners
\begin{equation}
C_{ai,-\nu}^{\rm Ed,\perp}=C_{ai,\nu}^{\rm Ed,\perp}.
\label{eq:SM_TR_Edelstein_even}
\end{equation}
This is the precise sense in which the ordinary dissipative Edelstein channel adds between time-reversed valleys.

\subsubsection{Electric-field Fermi-surface shift: the displacement is the same in both valleys}

It is useful to derive Eq.~\eqref{eq:SM_TR_Edelstein_even} directly from the shifted distribution, because the valley parity can otherwise be obscured by a radial shorthand such as $\mathrm d f/\mathrm d k$.  For equally occupied time-reversal partners the equilibrium occupation satisfies
\begin{equation}
f^0_{\eta,-\nu}(-\bq)=f^0_{\eta\nu}(\bq).
\label{eq:SM_TR_f0}
\end{equation}
A spatially uniform electric field produces the same displacement of the distribution in \emph{global} crystal momentum in the two valleys.  Denote that common displacement by $\delta\bk$.  To linear order,
\begin{equation}
\delta f_{\eta\nu}(\bq)
=-\delta k_i\,\partial_{q_i}f^0_{\eta\nu}(\bq).
\label{eq:SM_deltaf_shift}
\end{equation}
In the relaxation-time convention used throughout this work one may equivalently write
\begin{equation}
\delta f_{\eta\nu}(\bq)
=e\tau E_i v_{\eta\nu}^{i}(\bq)(-f'_{\eta\nu}),
\label{eq:SM_deltaf_boltzmann}
\end{equation}
with the overall sign fixed by the charge convention.  Equations~\eqref{eq:SM_deltaf_shift} and \eqref{eq:SM_deltaf_boltzmann} describe the same translation of the Fermi surface.  In particular, there is \emph{no} valley-dependent sign in $\delta\bk$.

The sign change appears instead when the same displacement acts on the slopes of the two time-reversed distributions.  Differentiating Eq.~\eqref{eq:SM_TR_f0} gives
\begin{equation}
\left[\partial_{q_i}f^0_{\eta,-\nu}(\bq)\right]_{\bq\to-\bq}
=-\partial_{q_i}f^0_{\eta\nu}(\bq),
\label{eq:SM_TR_grad_f0}
\end{equation}
and therefore, for the \emph{same} $\delta\bk$,
\begin{equation}
\boxed{\delta f_{\eta,-\nu}(-\bq)=-\delta f_{\eta\nu}(\bq).}
\label{eq:SM_TR_deltaf}
\end{equation}
The equivalent velocity statement follows immediately from
$v_{\eta,-\nu}^{i}(-\bq)=-v_{\eta\nu}^{i}(\bq)$.  Thus the field displaces both Fermi surfaces in the same global direction, while the occupation correction at a pair of time-reversed points has opposite sign because the equilibrium Fermi-surface normals are opposite.

This distinction is important for any response built from a time-reversal-odd band observable.  For real spin,
\begin{equation}
\langle s_a\rangle_{\eta,-\nu}(-\bq)\,\delta f_{\eta,-\nu}(-\bq)
=\langle s_a\rangle_{\eta\nu}(\bq)\,\delta f_{\eta\nu}(\bq),
\label{eq:SM_TR_spin_deltaf_even}
\end{equation}
which is the usual reason why a dissipative spin Edelstein response is allowed in a time-reversal-symmetric system.  The same logic applies to the parent OAM, for which
$L^a_{\eta,-\nu}(-\bq)=-L^a_{\eta\nu}(\bq)$.  The valley-resolved orbital Edelstein response may be written either as
\begin{align}
L_{{\rm OE},\nu}^{a}
&=\sum_\eta\int_{\bq}L^a_{\eta\nu}(\bq)\,\delta f_{\eta\nu}(\bq)
\nonumber\\
&=e\tau E_i\sum_\eta\int_{\bq}(-f'_{\eta\nu})
 v_{\eta\nu}^{i}(\bq)L^a_{\eta\nu}(\bq).
\label{eq:SM_LOE_deltaf}
\end{align}
Changing variables $\bq\to-\bq$ in the partner valley and using Eqs.~\eqref{eq:SM_TR_deltaf} and \eqref{eq:SM_TR_OAM} yields
\begin{align}
L_{{\rm OE},-\nu}^{a}
&=\sum_\eta\int_{\bq}L^a_{\eta,-\nu}(-\bq)\,\delta f_{\eta,-\nu}(-\bq)
\nonumber\\
&=\sum_\eta\int_{\bq}[-L^a_{\eta\nu}(\bq)]\,[-\delta f_{\eta\nu}(\bq)]
=L_{{\rm OE},\nu}^{a}.
\label{eq:SM_TR_LOE_from_shift}
\end{align}
Thus the \emph{equilibrium} OAM is valley odd, but its dissipative Fermi-surface response is valley even.  A notation such as $eE\tau\,\mathrm d f/\mathrm d k$ is safe only if the directional derivative is retained: the relevant object is $E_i\partial_{k_i}f^0$, or equivalently $E_i v_i f'$, not a scalar radial derivative with its angular sign suppressed.

A particularly important subclass is a velocity-like spin--orbit vertex
\begin{equation}
H_{\so,\nu}^{(v)}
=\hbar\Lambda_{jc,\nu}(\bq)v_{j,\nu}^{\orb}s_c,
\label{eq:SM_velocity_SOC_general}
\end{equation}
where time reversal requires $\Lambda_{jc,-\nu}(-\bq)=\Lambda_{jc,\nu}(\bq)$.  For a momentum-independent $\Lambda_{jc}$ this is automatic.  In this case
\begin{equation}
\beta_{\eta\nu c}=\Lambda_{jc,\nu}\partial_{q_j}\varepsilon_{\eta\nu},
\quad
\bb_{c\nu\perp}=d_\nu\Lambda_{jc,\nu}\partial_{q_j}\hat{\bd}_\nu,
\end{equation}
and Eq.~\eqref{eq:SM_generic_mixed_tensor} becomes
\begin{equation}
{\cal M}_{\eta\nu ic}
=\frac{2\Delta_{\eta\nu}}{\hbar}\Lambda_{jc,\nu}g_{ij,\nu},
\quad
\Delta_{\eta\nu}=\varepsilon_{\eta\nu}-\varepsilon_{\bar\eta\nu}=2\eta d_\nu,
\label{eq:SM_metric_reduction}
\end{equation}
with
\begin{equation}
g_{ij,\nu}=\frac14\partial_{q_i}\hat{\bd}_\nu\cdot
\partial_{q_j}\hat{\bd}_\nu.
\end{equation}
The metric is time-reversal even,
\begin{equation}
g_{ij,-\nu}(-\bq)=g_{ij,\nu}(\bq).
\label{eq:SM_TR_metric}
\end{equation}
Since Eq.~\eqref{eq:SM_velocity_SOC_general} is generally momentum dependent, it can also generate explicit $v_i^{\so}$ contributions to the full physical velocity; those remain separate from Eq.~\eqref{eq:SM_metric_reduction}.

The Berry curvature and the parent orbital moment instead sample the antisymmetric pseudospin texture,
\begin{align}
\Omega_{\eta\nu,ij}
&=-\frac{\eta}{2}\hat{\bd}_\nu\cdot
(\partial_{q_i}\hat{\bd}_\nu\times\partial_{q_j}\hat{\bd}_\nu),
\\
L_{\eta\nu}^{z}&=-\frac{\Delta_{\eta\nu}}{\hbar}\Omega_{\eta\nu,xy},
\label{eq:SM_two_band_OAM}
\end{align}
and therefore
\begin{equation}
\Omega_{\eta,-\nu,ij}(-\bq)=-\Omega_{\eta\nu,ij}(\bq),
\quad
L_{\eta,-\nu}^{z}(-\bq)=-L_{\eta\nu}^{z}(\bq).
\label{eq:SM_TR_OAM}
\end{equation}
Thus the OGSR controlled by the quantum metric and the equilibrium OAM have opposite valley parity in a time-reversal pair: the former adds in the longitudinal Edelstein response, while the latter cancels before any valley-odd occupation factor is introduced.

The same bookkeeping is useful for the intrinsic spin response.  For an exact band $N$ define the mixed spin--electric curvature
\begin{equation}
{\cal A}_{ai,N\nu}(\bq)
=2\operatorname{Im}\sum_{M\ne N}
\frac{\langle N\nu|s_a|M\nu\rangle
\langle M\nu|\partial_{q_i}H_\nu|N\nu\rangle}
{(E_{N\nu}-E_{M\nu})^2}.
\label{eq:SM_generic_spin_curvature_valley}
\end{equation}
Both $s_a$ and the velocity vertex are odd under time reversal, so their matrix-element product is complex conjugated without an additional minus sign.  Its imaginary part therefore changes sign:
\begin{equation}
{\cal A}_{ai,N,-\nu}(-\bq)
=-{\cal A}_{ai,N\nu}(\bq).
\label{eq:SM_TR_spin_curvature}
\end{equation}
Consequently a full intrinsic spin density of a time-reversal-symmetric, equally occupied valley pair cancels, whereas the dissipative Edelstein coefficient in Eq.~\eqref{eq:SM_TR_Edelstein_even} is valley even.  The tilted Dirac example below makes this distinction explicit term by term.

For reference, the valley parities established above may be summarized compactly as
\begin{equation}
\begin{array}{c|c}
X & X_{-\nu}(-\bq) \\
\hline
f^0_{\eta\nu},\; r_{\eta\nu} & +f^0_{\eta\nu},\;+r_{\eta\nu} \\
\partial_{q_i}f^0_{\eta\nu},\;v_{\eta\nu}^{i} & -\partial_{q_i}f^0_{\eta\nu},\;-v_{\eta\nu}^{i} \\
\delta f_{\eta\nu},\;\langle s_a\rangle_{\eta\nu} & -\delta f_{\eta\nu},\;-\langle s_a\rangle_{\eta\nu} \\
\hat{\bbeta}_{\eta\nu} & -\hat{\bbeta}_{\eta\nu} \\
{\cal M}_{\eta\nu ic},\; g_{ij,\nu} & +{\cal M}_{\eta\nu ic},\;+g_{ij,\nu} \\
\Omega_{\eta\nu,ij},\;L_{\eta\nu}^z
&-\Omega_{\eta\nu,ij},\;-L_{\eta\nu}^z \\
\langle s_a\rangle\delta f,\;L^z\delta f & +\langle s_a\rangle\delta f,\;+L^z\delta f \\
{\cal A}_{ai,N\nu}&-{\cal A}_{ai,N\nu}
\end{array}
\label{eq:SM_valley_parity_summary}
\end{equation}
Here every quantity on the right-hand side is evaluated at $(\nu,\bq)$.  In particular, the row containing $\delta f$ assumes the \emph{same} electric-field displacement $\delta\bk$ in the two valleys, as established in Eqs.~\eqref{eq:SM_deltaf_shift}--\eqref{eq:SM_TR_deltaf}.  The expectation values of real spin, velocity, and OAM are separately odd at paired points, while the dissipative products $\langle s_a\rangle\delta f$ and $L^z\delta f$ are even.  Equation~\eqref{eq:SM_valley_parity_summary} will be used below rather than assigning valley signs by inspection.

Finally, all formulas from Eq.~\eqref{eq:SM_beta_eta} onward assume valley conservation.  If the intervalley components $\tau_{x,y}^{\rm v}$ in Eq.~\eqref{eq:SM_full_valley_SOC} are present, the exact parent-projector formulation still applies after enlarging the parent Hilbert space to pseudospin $\otimes$ valley $\otimes$ real spin, but the simple two-band formulas no longer reduce the problem valley by valley.

\section{Radial two-band parent and quadratic Dirac deformation}
\label{sec:SM_radial_deformation}

The generic formulas above become especially transparent for a rotationally symmetric two-band parent.  We write
\begin{equation}
H_{p,\nu}=\epsilon_0(k)1
+d_\parallel(k)(\nu\cos\phi\,\sigma_x+\sin\phi\,\sigma_y)
+d_z(k)\sigma_z,
\label{eq:SM_radial_parent}
\end{equation}
where $\bq=k(\cos\phi,\sin\phi)$ is the local momentum and $\nu=\pm$ is kept explicitly.  For a time-reversal pair this parent obeys Eq.~\eqref{eq:SM_TR_parent}.  We choose the momentum-independent pseudospin Rashba vertex
\begin{equation}
H_{R,\nu}=\lambda_R(\nu\sigma_xs_y-\sigma_ys_x).
\label{eq:SM_fixed_pseudospin_Rashba}
\end{equation}
Equation~\eqref{eq:SM_fixed_pseudospin_Rashba} is a \emph{choice} of real-spin SOC appended to the pseudospin parent; it is not implied by Eq.~\eqref{eq:SM_radial_parent}.  Its advantage is that it is the same microscopic vertex used in the main text and has no explicit velocity contribution, $v_i^{\so}=0$.

For the parent conduction band let
\begin{equation}
d(k)=\sqrt{d_\parallel^2+d_z^2},
\quad
\varepsilon(k)=\epsilon_0(k)+d(k).
\end{equation}
Projection of Eq.~\eqref{eq:SM_fixed_pseudospin_Rashba} gives
\begin{align}
P_{+\nu}H_{R,\nu}P_{+\nu}
&=r(k)(\cos\phi\,s_y-\sin\phi\,s_x)P_{+\nu},
\\[-2pt]
r(k)&=\lambda_R\frac{d_\parallel(k)}{d(k)}.
\label{eq:SM_radial_r}
\end{align}
Hence, to first order in $\lambda_R$,
\begin{equation}
E_{\nu\zeta}(k)=\varepsilon(k)+\zeta r(k),
\quad
\langle s_y\rangle_{\nu\zeta}=\zeta\cos\phi.
\label{eq:SM_radial_branches}
\end{equation}
Because the Rashba vertex is momentum independent, the physical velocity is purely orbital.  The parent-diagonal and parent-off-diagonal pieces of its branch expectation are then
\begin{align}
\langle v_{x,\nu}^{\parallel}\rangle_{\nu\zeta}
&=\frac{\varepsilon'(k)}{\hbar}\cos\phi+O(\lambda_R^2),
\\
\langle v_{x,\nu}^{\perp}\rangle_{\nu\zeta}
&=\zeta\frac{r'(k)}{\hbar}\cos\phi+O(\lambda_R^2).
\label{eq:SM_radial_vsplit}
\end{align}
The second equation may also be obtained directly from Eq.~\eqref{eq:SM_generic_vperp}.  It has a simple interpretation: the parent-off-diagonal orbital velocity is the part of the Rashba-induced branch velocity that is missed if one transports the carriers using only the projected parent-band group velocity.

Expanding the two Fermi contours about the parent radius $k_F$, defined by $\mu=\varepsilon(k_F)$ and assuming $\varepsilon'(k_F)>0$, gives a closed weak-Rashba decomposition for valley $\nu$,
\begin{align}
C_{yx,\nu}^{\parallel}
&=-\frac{1}{2\pi\hbar}
\left.\frac{r+kr'}{\varepsilon'}\right|_{k_F},
\label{eq:SM_radial_Cparallel}
\\
C_{yx,\nu}^{\perp}
&=\frac{1}{2\pi\hbar}
\left.\frac{kr'}{\varepsilon'}\right|_{k_F},
\label{eq:SM_radial_Cperp}
\\
C_{yx,\nu}
&=-\frac{1}{2\pi\hbar}
\left.\frac{r}{\varepsilon'}\right|_{k_F}.
\label{eq:SM_radial_Ctotal}
\end{align}
The derivation uses only $\int_0^{2\pi}\cos^2\phi\,\dd\phi=\pi$ and the first-order contour displacement.  Equation~\eqref{eq:SM_radial_Cperp} shows that counterflow occurs whenever the projected Rashba splitting grows with $k$ on an electron-like Fermi contour, $r'(k_F)>0$.  Its sign is therefore not imposed by projector algebra alone, in agreement with the general discussion in the main text.
For a time-reversal pair the projected splitting $r(k)$ is valley independent, and Eqs.~\eqref{eq:SM_radial_Cparallel}--\eqref{eq:SM_radial_Ctotal} therefore obey
\begin{equation}
C_{yx,-\nu}^{\parallel}=C_{yx,\nu}^{\parallel},\quad
C_{yx,-\nu}^{\perp}=C_{yx,\nu}^{\perp},\quad
C_{yx,-\nu}=C_{yx,\nu}.
\label{eq:SM_radial_valley_even}
\end{equation}
For two equally occupied valleys the physical longitudinal coefficient is twice the value quoted for one valley.

We now deform the massive Dirac parent without changing the pseudospin Rashba vertex,
\begin{equation}
H_{p,\nu}=Dk^2 1
+\alpha(\nu k_x\sigma_x+k_y\sigma_y)
+(m+Bk^2)\sigma_z.
\label{eq:SM_deformed_Dirac}
\end{equation}
The two additional terms have distinct roles: $Dk^2$ produces particle--hole asymmetry, while $Bk^2\sigma_z$ bends the mass texture.  Define
\begin{align}
M(k)&=m+Bk^2,
\\
d(k)&=\sqrt{\alpha^2k^2+M(k)^2},
\\
{\cal V}(k)&=2D+\frac{\alpha^2+2BM(k)}{d(k)},
\\
\varepsilon'(k)&=k{\cal V}(k).
\end{align}
The projected Rashba splitting and its derivative are
\begin{align}
r(k)&=\lambda_R\frac{\alpha k}{d(k)},
\\
r'(k)&=\lambda_R\alpha
\frac{M(k)[m-Bk^2]}{d(k)^3}.
\label{eq:SM_deformed_rprime}
\end{align}
Substitution in Eqs.~\eqref{eq:SM_radial_Cparallel}--\eqref{eq:SM_radial_Ctotal} gives, at the parent Fermi radius $k_F$,
\begin{align}
C_{yx,\nu}^{\parallel}
&=-\frac{\lambda_R\alpha}
{2\pi\hbar\,d_F{\cal V}_F}
\left[1+\frac{M_F(m-Bk_F^2)}{d_F^2}\right],
\label{eq:SM_deformed_Cparallel}
\\
C_{yx,\nu}^{\perp}
&=\frac{\lambda_R\alpha}
{2\pi\hbar\,d_F{\cal V}_F}
\frac{M_F(m-Bk_F^2)}{d_F^2},
\label{eq:SM_deformed_Cperp}
\\
C_{yx,\nu}
&=-\frac{\lambda_R\alpha}
{2\pi\hbar\,d_F{\cal V}_F},
\label{eq:SM_deformed_Ctotal}
\end{align}
where $M_F=M(k_F)$, $d_F=d(k_F)$, and ${\cal V}_F={\cal V}(k_F)$.  The corresponding counterflow fraction is
\begin{equation}
{\cal B}_{yx}
\equiv\frac{C_{yx,\nu}^{\perp}}{|C_{yx,\nu}^{\parallel}|}
=\frac{A_F}{1+A_F},
\quad
A_F=\frac{M_F(m-Bk_F^2)}{d_F^2},
\label{eq:SM_deformed_fraction}
\end{equation}
whenever $A_F>0$.  The $D$ term changes the absolute response through the Fermi velocity and, at fixed chemical potential, through $k_F$, but it does not create or remove the counterflow by itself.  The $B$ term changes the pseudospin texture and can eventually reverse the sign when $M_F(m-Bk_F^2)$ changes sign.  Thus the opposing contribution is robust over a finite region around the massive-Dirac point, while its sign is not universal.

This deformation also illustrates why the quantum-metric language must be tied to a specified spin--orbit vertex.  The radial metric of Eq.~\eqref{eq:SM_deformed_Dirac} is
\begin{equation}
g_{kk}=\frac{\alpha^2(m-Bk^2)^2}{4d(k)^4}.
\label{eq:SM_deformed_metric}
\end{equation}
The corresponding conduction-band Berry curvature and OAM keep the opposite valley parity,
\begin{align}
\Omega_{+\nu,xy}(k)
&=-\nu\frac{\alpha^2(m-Bk^2)}{2d(k)^3},\\
L_{+\nu}^{z}(k)
&=\nu\frac{\alpha^2(m-Bk^2)}{\hbar d(k)^2}.
\label{eq:SM_deformed_OAM_valley}
\end{align}
Thus the longitudinal counterflow in Eq.~\eqref{eq:SM_deformed_Cperp} is valley even even though the parent OAM in the same band is valley odd.
For $B\neq0$, Eq.~\eqref{eq:SM_deformed_Cperp} is controlled by the mixed tensor of Eq.~\eqref{eq:SM_generic_mixed_tensor} and is not simply proportional to $g_{kk}$, because the fixed pseudospin Rashba vertex in Eq.~\eqref{eq:SM_fixed_pseudospin_Rashba} is no longer proportional to the full deformed parent velocity.  If instead the velocity-like SOC in Eq.~\eqref{eq:SM_velocity_SOC_general} is chosen, the reduction to the metric is exact by Eq.~\eqref{eq:SM_metric_reduction}.

\subsection{Linear massive-Dirac limit}

Setting $B=D=0$ gives $d=\varepsilon_k=(\alpha^2k^2+m^2)^{1/2}$ and ${\cal V}=\alpha^2/\varepsilon_k$.  Equations~\eqref{eq:SM_deformed_Cparallel}--\eqref{eq:SM_deformed_Ctotal} therefore reduce, in either valley, to
\begin{align}
C_{yx,\nu}^{\parallel}
&=-\frac{\lambda_R}{2\pi\hbar\alpha}
\left(1+\frac{m^2}{\mu^2}\right),
\label{eq:SM_Cparallel}
\\
C_{yx,\nu}^{\perp}
&=\frac{\lambda_R}{2\pi\hbar\alpha}
\frac{m^2}{\mu^2},
\label{eq:SM_Cperp}
\\
C_{yx,\nu}&=-\frac{\lambda_R}{2\pi\hbar\alpha},
\quad
{\cal B}_{yx}=\frac{m^2}{\mu^2+m^2}.
\label{eq:SM_linear_Dirac_limit}
\end{align}
The complete cancellation of the mass-dependent correction in the total coefficient is therefore a special simplification of the linear Dirac dispersion.  The existence of an opposing parent-off-diagonal contribution is not: it follows from the more general radial identity in Eq.~\eqref{eq:SM_radial_Cperp} and remains finite under the quadratic deformation in Eq.~\eqref{eq:SM_deformed_Dirac}.

\section{Modern-theory OAM response}

For reference, this section records the complete OAM response used in the main text.  We follow the notation of the companion modern-theory OME work cited there and write
\begin{equation}
\begin{aligned}
L_E&=L_{\rm OE}+L_{E1}+L_{E2}+L_{E3},\\
L_{\rm OE}^a&=e\tau E_i\sum_n\int_{\bk}(-f_n')v_n^iL_n^a.
\end{aligned}
\label{eq:SM_LE_decomposition}
\end{equation}
The remaining terms are quoted directly rather than rederived from the density matrix.  All Berry connections in the following expressions are interband, ${\cal R}_{nm}^a=\hbar v_{nm}^a/[i(E_n-E_m)]$ for $n\ne m$, and $D_i{\cal R}^a=\partial_{k_i}{\cal R}^a-i[\mathcal R_i,{\cal R}^a]$.

{\small
\begin{subequations}
\begin{align}
L_{E1}^a={}&-eE_i\epsilon_{abc}
\sum_{n\ne m}\int_{\bk}\frac{f_n}{E_n-E_m}
\nonumber\\[-2pt]
&\times\operatorname{Re}\Tr\!\left[
(v_{nn}^b{\cal R}_{nm}^c+{\cal R}_{nm}^cv_{mm}^b)
{\cal R}_{mn}^i\right],
\label{eq:SM_L1_general}
\\
L_{E2}^a={}&-\frac{eE_i}{2\hbar}\epsilon_{abc}
\sum_n\int_{\bk}f_n
\Tr\!\left[\Pi_n\{{\cal R}^c,D_i{\cal R}^b\}\Pi_n\right]
\nonumber\\[-2pt]
&+eE_i\epsilon_{abc}\sum_{n\ne m}\int_{\bk}
\frac{f_n}{E_n-E_m}
\nonumber\\[-2pt]
&\times\operatorname{Re}\Tr\!\left[
(v_{nn}^c{\cal R}_{nm}^b+{\cal R}_{nm}^bv_{mm}^c)
{\cal R}_{mn}^i\right],
\label{eq:SM_L2_general}
\\
L_{E3}^a={}&-\frac{eE_i}{2}\epsilon_{abc}
\sum_{n,m,l}'\int_{\bk}\frac{f_n}{E_m-E_n}
\nonumber\\[-2pt]
&\times\Tr\!\left[
 {\cal R}_{nm}^i{\cal R}_{ml}^bv_{ln}^c
+{\cal R}_{nl}^bv_{lm}^c{\cal R}_{mn}^i
\right.\nonumber\\[-2pt]
&\hspace{8mm}\left.
+v_{nl}^c{\cal R}_{lm}^b{\cal R}_{mn}^i
+{\cal R}_{nm}^iv_{ml}^c{\cal R}_{ln}^b\right].
\label{eq:SM_L3_general}
\end{align}
\end{subequations}
}

The first term, $L_{E1}$, contains the velocity-sum structure.  The complete itinerant-circulation term $L_{E2}$ contains both the connection--derivative tensor exhibited in Eq.~\eqref{eq:SM_T_tensor} and a second velocity-sum term.  The local-circulation contribution $L_{E3}$ contains genuine three-manifold paths and therefore vanishes for a two-band parent model.  Resolving each velocity and connection into the four parent blocks of Eq.~\eqref{eq:SM_total_four_blocks} gives the exact parent-projector classification used in the main text.

The valley parity of these terms follows directly from time reversal and is useful independently of the Dirac model.  Choose a time-reversal-related gauge in which the same band label denotes paired states in the two valleys; in a general gauge the same statements hold with the usual sewing matrices, which cancel from the response traces.  At paired local momenta,
\begin{align}
v_{nm,-\nu}^{a}(-\bq)
&=-[v_{nm,\nu}^{a}(\bq)]^*,
\\
{\cal R}_{nm,-\nu}^{a}(-\bq)
&=[{\cal R}_{nm,\nu}^{a}(\bq)]^*,
\\
[D_i{\cal R}^{a}]_{nm,-\nu}(-\bq)
&=-[D_i{\cal R}^{a}]_{nm,\nu}(\bq)^*.
\label{eq:SM_TR_OAM_building_blocks}
\end{align}
The first relation expresses the odd parity of velocity, the second follows from ${\cal R}_{nm}^{a}=\hbar v_{nm}^{a}/[i(E_n-E_m)]$, and the third includes the sign from differentiating at $-\bq$.  Substitution into Eqs.~\eqref{eq:SM_L1_general}--\eqref{eq:SM_L3_general} shows, for equally occupied time-reversal partners,
\begin{equation}
L_{j,-\nu}^{a}=-L_{j,\nu}^{a},
\quad j=1,2,3.
\label{eq:SM_TR_intrinsic_L_general}
\end{equation}
The orbital Edelstein term has the opposite parity.  Using Eq.~\eqref{eq:SM_LOE_deltaf}, its Fermi-surface integrand is the product of two time-reversal-odd quantities, $v_n^iL_n^a$, or equivalently $L_n^a\delta f_n$.  The field-induced momentum displacement itself is common to the two valleys; Eq.~\eqref{eq:SM_TR_deltaf} shows that the occupation correction at paired points is nevertheless odd because the equilibrium gradients are opposite.  Hence
\begin{equation}
L_{{\rm OE},-\nu}^{a}=L_{{\rm OE},\nu}^{a}.
\label{eq:SM_TR_LOE_general}
\end{equation}
Thus a time-reversal-invariant system may have an additive dissipative orbital Edelstein response while the intrinsic $L_{E1}+L_{E2}+L_{E3}$ contribution cancels between a pair of equally occupied valleys.  The explicit tilted-Dirac formulas below provide a direct realization of these general parities.

\section{Exact massive-Dirac spectrum}

The full tilted model used in the Letter is
\begin{align}
H_\nu(\bq)={}&\hbar v_{t,\nu}q_x
+\alpha(\nu q_x\sigma_x+q_y\sigma_y)+m\sigma_z
\nonumber\\
&+\lambda_R(\nu\sigma_xs_y-\sigma_ys_x),
\quad v_{t,\nu}=\nu v_t.
\label{eq:SM_full_tilted_Dirac}
\end{align}
The scalar tilt changes the energies and Fermi contours but not the
eigenspinors at fixed momentum.  For the untilted spectrum, rotational
invariance allows the local momentum to be chosen as $\bq=k\hat{\bm x}$.
Keeping the valley index explicitly,
\begin{equation}
H_\nu(k\hat{\bm x})=\alpha\nu k\sigma_x+m\sigma_z
+\lambda_R(\nu\sigma_xs_y-\sigma_ys_x).
\label{eq:SM_Hnu_x}
\end{equation}
The square is
\begin{equation}
H_\nu^2=(\alpha^2k^2+m^2+2\lambda_R^2)1
+2\alpha k\lambda_Rs_y
-2\nu\lambda_R^2\sigma_zs_z.
\label{eq:SM_H2}
\end{equation}
The two matrices in the last line anticommute and square to the identity.  The explicit sign $\nu$ therefore drops out of the eigenvalues of $H_\nu^2$, as required by time reversal.  The positive energies are
\begin{equation}
E_s^2
=\alpha^2k^2+m^2+2\lambda_R^2
+2s\lambda_R\sqrt{\alpha^2k^2+\lambda_R^2},
\end{equation}
or, equivalently,
\begin{equation}
E_s^2=m^2+\left(\sqrt{\alpha^2k^2+\lambda_R^2}+s\lambda_R\right)^2.
\end{equation}
Particle--hole symmetry supplies the two signs $E_{\rho s,\nu}=\rho E_s$; the spectrum is identical in the two valleys.

For electron doping, $p=(\mu^2-m^2)^{1/2}$ and the equation $E_s(k_s)=\mu$ gives
\begin{equation}
\sqrt{\alpha^2k_s^2+\lambda_R^2}+s\lambda_R=p,
\end{equation}
hence
\begin{equation}
k_s^2=\frac{p(p-2s\lambda_R)}{\alpha^2}.
\end{equation}
The upper contour $s=+$ disappears when $p=2\lambda_R$.

\section{Massive-Dirac band velocities}

For the momentum-independent Rashba interaction, $v_i^{\so,\parallel}=v_i^{\so,\perp}=0$. In this and the following massive-Dirac sections only, $v_i^\parallel$ and $v_i^\perp$ abbreviate the two parent-orbital blocks.

For the two ordinary parent bands it is convenient to use the rank-one orbital matrices
\begin{align}
P_{\eta\nu}&=\frac{1}{2}(1+\eta\hat{\bd}_\nu\cdot\bsigma),
\\
\hat{\bd}_\nu&=\frac{(\alpha\nu q_x,\alpha q_y,m)}{\varepsilon_k},
\quad k=|\bq|.
\end{align}
Here $P_{\eta\nu}=|u_{\eta\nu}\rangle\langle u_{\eta\nu}|$ is only an algebraic shorthand for the conduction- or valence-band spinor. No composite-band or non-Abelian construction is required in the massive-Dirac calculation.
For any Pauli matrix $\sigma_a$,
\begin{equation}
\sum_\eta P_{\eta\nu}\sigma_aP_{\eta\nu}
=\hat d_{\nu a}(\hat{\bd}_\nu\cdot\bsigma),
\end{equation}
and therefore
\begin{align}
v_{x,\nu}^\parallel
&=\frac{\alpha\nu}{\hbar}\hat d_{\nu x}(\hat{\bd}_\nu\cdot\bsigma),
&
v_{x,\nu}^\perp
&=\frac{\alpha\nu}{\hbar}
[\sigma_x-\hat d_{\nu x}(\hat{\bd}_\nu\cdot\bsigma)],
\\
v_{y,\nu}^\parallel
&=\frac{\alpha}{\hbar}\hat d_{\nu y}(\hat{\bd}_\nu\cdot\bsigma),
&
v_{y,\nu}^\perp
&=\frac{\alpha}{\hbar}
[\sigma_y-\hat d_{\nu y}(\hat{\bd}_\nu\cdot\bsigma)].
\end{align}
Direct substitution verifies $v_{i,\nu}^\parallel+v_{i,\nu}^\perp=v_{i,\nu}^{\orb}$.

At zero temperature the two-dimensional Fermi-surface integral is
\begin{align}
\int\frac{\dd^2q}{(2\pi)^2}
\delta(\mu-E_s)F_s(k,\phi)
=\frac{k_s}{(2\pi)^2}
\int_0^{2\pi}\dd\phi\,
\frac{F_s(k_s,\phi)}{|\partial_kE_s(k_s)|}.
\label{eq:SM_FS_integral}
\end{align}
The numerical calculation diagonalizes the Hermitian $4\times4$ Hamiltonian at each $(k_s,\phi)$ and evaluates
\begin{align}
C_{yx,\nu}^\parallel
&=\sum_s\frac{k_s}{(2\pi)^2}
\int\dd\phi\,
\frac{\langle s_y\rangle_{s\nu}\langle v_{x,\nu}^\parallel\rangle_{s\nu}}
{|\partial_kE_s|},
\\
C_{yx,\nu}^\perp
&=\sum_s\frac{k_s}{(2\pi)^2}
\int\dd\phi\,
\frac{\langle s_y\rangle_{s\nu}\langle v_{x,\nu}^\perp\rangle_{s\nu}}
{|\partial_kE_s|}.
\end{align}
For every point, the numerical identity
\begin{equation}
C_{yx,\nu}=C_{yx,\nu}^\parallel+C_{yx,\nu}^\perp
\end{equation}
is satisfied to machine precision before plotting.  The time-reversal partner gives the same $E_x\to s_y$ coefficient, in agreement with Eq.~\eqref{eq:SM_radial_valley_even}.

\section{OAM identity and Feshbach self-energy}

For a nondegenerate parent band $P_n$ the band OAM may be written
\begin{equation}
L_n^a=\epsilon_{aij}\sum_{m\neq n}\operatorname{Re}
\left[{\cal R}_{nm}^{i}v_{mn}^{j}\right].
\label{eq:SM_L_def}
\end{equation}
With
\begin{equation}
v_{nm}^{i}=\frac{i}{\hbar}(\varepsilon_n-\varepsilon_m)
{\cal R}_{nm}^{i},
\end{equation}
one obtains
\begin{equation}
L_n^aP_n=-\frac{\hbar}{2}\epsilon_{aij}
\sum_{m\neq n}\frac{i\left(
P_nv_iP_mv_jP_n-P_nv_jP_mv_iP_n
\right)}{\varepsilon_n-\varepsilon_m}.
\label{eq:SM_L_identity}
\end{equation}
This is the multiband form of the equilibrium parent-band OAM. For the two-band Dirac application, $P$ and $Q$ below denote the ordinary conduction and valence bands within a fixed valley $\nu$. Its $z$ component reduces to
\begin{equation}
\frac{i}{\varepsilon_P-\varepsilon_Q}
\left(Pv_{x,\nu}Qv_{y,\nu}P-Pv_{y,\nu}Qv_{x,\nu}P\right)
=-\frac{1}{\hbar}L_{P,\nu}^zP.
\label{eq:SM_L_identity_two_band}
\end{equation}

For the Dirac Rashba vertex in valley $\nu$,
\begin{equation}
H_{R,\nu}=\frac{\hbar\lambda_R}{\alpha}
(v_{x,\nu}s_y-v_{y,\nu}s_x),
\end{equation}
and Pauli-matrix multiplication gives
{\small
\begin{align}
PH_{R,\nu}QH_{R,\nu}P
=&\left(\frac{\hbar\lambda_R}{\alpha}\right)^2
\Big[Pv_{x,\nu}Qv_{x,\nu}P
\nonumber\\[-2pt]
&+Pv_{y,\nu}Qv_{y,\nu}P
+iPv_{x,\nu}Qv_{y,\nu}Ps_z
\nonumber\\[-2pt]
&-iPv_{y,\nu}Qv_{x,\nu}Ps_z\Big].
\end{align}
}
Using Eq.~\eqref{eq:SM_L_identity_two_band},
\begin{equation}
\frac{PH_{R,\nu}QH_{R,\nu}P}{\varepsilon_P-\varepsilon_Q}
=C_{P,\nu}P-\frac{\hbar\lambda_R^2}{\alpha^2}L_{P,\nu}^zs_z,
\end{equation}
where
\begin{equation}
C_{P,\nu}P=\left(\frac{\hbar\lambda_R}{\alpha}\right)^2
\frac{Pv_{x,\nu}Qv_{x,\nu}P+Pv_{y,\nu}Qv_{y,\nu}P}{\varepsilon_P-\varepsilon_Q}
\end{equation}
is the symmetric scalar sector.

For completeness, the exact elimination of $Q$ follows directly from the block Schr\"odinger equation. Write $|p\rangle=P|\Psi\rangle$ and $|q\rangle=Q|\Psi\rangle$. The equations $P(H-E)|\Psi\rangle=0$ and $Q(H-E)|\Psi\rangle=0$ are
\begin{align}
(PHP-E)|p\rangle+PHQ|q\rangle&=0,
\nonumber\\
QHP|p\rangle+(QHQ-E)|q\rangle&=0.
\end{align}
When $E-QHQ$ is invertible, the second equation gives
\begin{equation}
|q\rangle=(E-QHQ)^{-1}QHP|p\rangle.
\end{equation}
Substituting this result into the first equation proves
\begin{equation}
H_{\rm eff}^{P}(E)=PHP+PHQ(E-QHQ)^{-1}QHP.
\label{eq:SM_Feshbach}
\end{equation}
The identity is exact but energy dependent. It sums repeated propagation inside the full $Q$ block to all orders; it becomes an ordinary energy-independent L\"owdin Hamiltonian only after a controlled expansion of the resolvent about the parent-band energy. At a pole of the $Q$ resolvent the resonant state cannot be eliminated and must instead be retained in the active subspace.

At strong Rashba coupling the resolvent can be expanded in the spin basis,
\begin{equation}
Q(E-QHQ)^{-1}Q=R_0(E)Q+\sum_aR_a(E)Qs_aQ.
\end{equation}
The self-energy then contains the symmetric and antisymmetric velocity products above together with additional spin contractions. The coefficient multiplying the antisymmetric product is the nonperturbative continuation of the OAM-related sector, but it is no longer described by a single energy-independent $L_{P,
u}^zs_z$ coefficient.

\section{OAM-resolved massive-Dirac response}

The exact counterflow calculated in the main text is the leading $\tau$-linear Fermi-surface response. At first order in Rashba coupling its parent-off-diagonal term contains the symmetric two-connection tensor. It is the OGSR, but it is not the Fermi-surface shift of the antisymmetric equilibrium moment and hence is not $L_{\rm OE}$. Nor should it be identified with the intrinsic $L_{E1},L_{E2},L_{E3}$ contributions quoted in the companion OME paper.

At second order define
\begin{align}
B_{\eta\nu}&=\operatorname{Im}
\left({\cal R}_{\eta\nu,\bar\eta\nu}^{x}
{\cal R}_{\bar\eta\nu,\eta\nu}^{y}\right),
&
\Delta_{\eta\nu}&=\varepsilon_{\eta\nu}-\varepsilon_{\bar\eta\nu},
\\
L_{\eta\nu}^z&=+\frac{2\Delta_{\eta\nu}}{\hbar}B_{\eta\nu},
&
r_k&=\frac{\lambda_R\alpha k}{\varepsilon_k},
\\
d_{\eta\nu}^{(z)}&=-\frac{\lambda_R^2\hbar}{\alpha^2}L_{\eta\nu}^z,
&
F_{\eta\nu}^{(q)}&=-\frac{\partial^q f_{\eta\nu}}{\partial\varepsilon_{\eta\nu}^q}.
\end{align}
The complete band-diagonal second-order kernel is
\begin{align}
S_{\rm FS}^{a,(2)}
=e\tau E_i\sum_{\eta,\nu}\int_{\bq}
\bigg[&
8\frac{\lambda_R^2}{\alpha^2}
F_{\eta\nu}^{(1)}v_{\eta\nu}^iB_{\eta\nu}
+2F_{\eta\nu}^{(1)}\frac{d_{\eta\nu}^{(z)}}{r_k}
\frac{\partial_{q_i}r_k}{\hbar}
\nonumber\\
&+2F_{\eta\nu}^{(2)}d_{\eta\nu}^{(z)}v_{\eta\nu}^i
\bigg]\delta_{az}.
\label{eq:SM_complete_second_spin_kernel}
\end{align}
All three terms descend from the antisymmetric loop and give the spin response associated with the OEE. The factor $\delta_{az}$ shows that this quadratic sector produces only $s_z$, not a correction to $E_x\to s_y$; it vanishes in the untilted model because every integrand is odd in $q_i$. The additional inverse gap, $\partial_{q_i}r_k$, and $f''$ factors arise from the spin-orbit vertex.

\subsection{Tilted cone: finite OEE contribution to spin}

Add $u_\nu q_x1$, with $u_\nu=\hbar v_{t,\nu}$, to valley $\nu$.  We keep $v_{t,\nu}$ arbitrary while deriving the valley-resolved response.  If $\nu=\pm$ are time-reversal partners, Eq.~\eqref{eq:SM_TR_condition} requires
\begin{equation}
v_{t,-\nu}=-v_{t,\nu},
\quad\text{or equivalently}\quad
v_{t,\nu}=\nu v_t
\label{eq:SM_TR_tilt}
\end{equation}
for a common positive parameter $v_t$.  The projectors, interband connections and moment are unchanged by the scalar tilt, while
\begin{equation}
v_{+,x,\nu}=v_{t,\nu}+\frac{\alpha^2q_x}{\hbar\varepsilon_k},
\quad
L_{+\nu}^{z}=+\frac{\nu m\alpha^2}{\hbar\varepsilon_k^2}.
\label{eq:SM_tilt_velocity_moment}
\end{equation}
The exact conduction-band energies are $E_s(k)=\mathcal E_s(k)$ in the notation of the main text. Put
\begin{equation}
\chi_k=(\alpha^2k^2+\lambda_R^2)^{1/2},
\quad
\langle s_z\rangle_{s\nu}=-\frac{s\nu m\lambda_R}{E_s\chi_k}.
\label{eq:SM_exact_sz}
\end{equation}
Using Eq.~\eqref{eq:SM_tilt_velocity_moment}, the exact spin texture factorizes as
\begin{equation}
\langle s_z\rangle_{s\nu}
=-\frac{s\hbar\lambda_R\varepsilon_k^2}
{\alpha^2E_s\chi_k}L_{+\nu}^{z}(k).
\label{eq:SM_exact_sz_L}
\end{equation}
This is an operator-derived conversion vertex, not a ratio defined after integration. It proves that the complete $E_x\!\to S_z$ Fermi-surface response of the tilted model is OAM-derived at every $k$ and to all orders in $\lambda_R$.

The closed Fermi-surface result can be obtained without expanding in Rashba coupling. For any radial band observable $g_s(k)$, expanding only to first order in the scalar tilt gives
\begin{align}
\frac{C_{g,\nu}}{v_{t,\nu}}
&=\sum_{s\in{\rm FS}}\left[
\frac{k_sg_s}{2\pi\partial_kE_s}
-\frac{1}{4\pi}\frac{\partial}{\partial\mu}
(k_s^2g_s)\right]
\nonumber\\
&=-\sum_{s\in{\rm FS}}
\frac{k_s^2}{4\pi}\frac{\partial g_s(k_s)}{\partial\mu}.
\label{eq:SM_tilt_radial_identity}
\end{align}
The cancellation in the second line uses $\partial_\mu k_s^2=2k_s/\partial_kE_s$. It is valid separately for every monotonic Fermi contour. With $p=(\mu^2-m^2)^{1/2}$,
\begin{equation}
\chi_s=p-s\lambda_R,
\quad
k_s^2=\frac{p(p-2s\lambda_R)}{\alpha^2},
\end{equation}
and $s=+$ is retained only for $p>2\lambda_R$.

Substitution of $g_s=\langle s_z\rangle_{s\nu}$ into Eq.~\eqref{eq:SM_tilt_radial_identity} gives
{\small
\begin{align}
C_{zx,\nu}^{[L]}
=-&\frac{\nu v_{t,\nu}m\lambda_R}{4\pi\alpha^2}
\sum_{s\in{\rm FS}}s\,p(p-2s\lambda_R)
\nonumber\\[-2pt]
&\times\left[
\frac{1}{\mu^2(p-s\lambda_R)}
+\frac{1}{p(p-s\lambda_R)^2}
\right]+O(v_{t,\nu}^3).
\label{eq:SM_tilt_spin_all_orders}
\end{align}
}
When both contours exist, their terms linear in $\lambda_R$ cancel and the exact sum reduces to
\begin{equation}
C_{zx,\nu}^{[L]}
=\frac{\nu v_{t,\nu}m\lambda_R^2}{2\pi\alpha^2}
\left[
\frac{p^2}{\mu^2(p^2-\lambda_R^2)}
+\frac{2\lambda_R^2}{(p^2-\lambda_R^2)^2}
\right].
\label{eq:SM_tilt_spin_two_contours}
\end{equation}
Beyond $p=2\lambda_R$ only the $s=-$ term of Eq.~\eqref{eq:SM_tilt_spin_all_orders} remains. This is why the strong-coupling response is not captured by continuing the $O(\lambda_R^2)$ formula.

Applying Eq.~\eqref{eq:SM_tilt_radial_identity} instead to the parent moment gives the exact OEE of the Rashba-split bands,
{\small
\begin{align}
\frac{L_{{\rm OE},x,\nu}^{z,{\rm par}}}{e\tau E_x}
=&\frac{\nu v_{t,\nu}m\mu}{2\pi\hbar}
\sum_{s\in{\rm FS}}
\frac{(p-2s\lambda_R)(p-s\lambda_R)}
{(\mu^2-2s\lambda_Rp)^2}
+O(v_{t,\nu}^3).
\label{eq:SM_tilt_LOE_all_orders}
\end{align}
}
Therefore
\begin{equation}
\eta_{{\rm OE},\nu}^{\rm exact}
=\frac{C_{zx,\nu}^{[L]}}
{L_{{\rm OE},x,\nu}^{z,{\rm par}}/(e\tau E_x)}
\label{eq:SM_tilt_eta_exact}
\end{equation}
is known explicitly to all orders in $\lambda_R$ whenever the parent-OAM response in the denominator is nonzero. For $m,\mu>0$ in the weak two-contour regime, $\eta_{{\rm OE},\nu}>0$, so $s_z$ follows $L_{\rm OE}^z$ and no independent $s_z$ term competes. At the band edge $\mu=m$ both responses vanish, so no ratio is defined. In the two-contour weak-coupling limit the spin response becomes
\begin{equation}
C_{zx,\nu}^{[L]}=\frac{\nu v_{t,\nu}m\lambda_R^2}
{2\pi\alpha^2\mu^2}+O(v_t\lambda_R^4).
\label{eq:SM_tilt_weak_limit}
\end{equation}
The code evaluates the finite-tilt Fermi contours directly and verifies Eq.~\eqref{eq:SM_tilt_spin_all_orders}; the perturbative expression is used only as the limiting check.
Both the spin response in Eqs.~\eqref{eq:SM_tilt_spin_all_orders}--\eqref{eq:SM_tilt_weak_limit} and the orbital Edelstein response in Eq.~\eqref{eq:SM_tilt_LOE_all_orders} are proportional to $\nu v_{t,\nu}$.  Hence a time-reversal pair with $v_{t,\nu}=\nu v_t$ gives equal contributions from the two valleys:
\begin{equation}
C_{zx,-\nu}^{[L]}=C_{zx,\nu}^{[L]},
\quad
L_{{\rm OE},x,-\nu}^{z,{\rm par}}
=L_{{\rm OE},x,\nu}^{z,{\rm par}}.
\label{eq:SM_TR_tilted_FS_add}
\end{equation}
The valley-odd equilibrium moment is compensated by the valley-odd \emph{equilibrium band tilt}, so the net Fermi-surface OAM-to-spin response survives the valley sum.  The tilt parameter $v_{t,\nu}$ should not be confused with the electric-field displacement of the Fermi surface: time reversal requires $v_{t,-\nu}=-v_{t,\nu}$ because the scalar term $\hbar v_{t,\nu}q_x$ belongs to the equilibrium Hamiltonian, whereas the nonequilibrium displacement $\delta\bk$ generated by the same applied field is the same global vector in both valleys.  The latter nevertheless produces the odd occupation correction in Eq.~\eqref{eq:SM_TR_deltaf} at paired points.

For the orthogonal field $E_y$, the spinless two-band intrinsic OME is
\begin{align}
L_{E1,\nu}^{z}
&=\frac{eE_yv_{t,\nu}(m^2+3\mu^2)}{48\pi\mu^3},
\nonumber\\
L_{E2,\nu}^{z}
&=\frac{eE_yv_{t,\nu}m^2}{12\pi\mu^3},
\quad L_{E3,\nu}^{z}=0,
\nonumber\\
L_{E,\rm int,\nu}^{z}
&=\frac{eE_yv_{t,\nu}(5m^2+3\mu^2)}{48\pi\mu^3}
\quad(\mu>m).
\label{eq:SM_tilt_intrinsic_OME}
\end{align}
In the gap $L_{E1,\nu}^{z}=L_{E2,\nu}^z=eE_yv_{t,\nu}/(12\pi m)$, $L_{E3,\nu}^z=0$, and $L_{E,\rm int,\nu}^{z}=eE_yv_{t,\nu}/(6\pi m)$. Because a scalar tilt leaves the eigenspinors, interband gaps, and insulating occupations unchanged, the corresponding intrinsic spin coherence vanishes.  Unlike the Fermi-surface OEE above, $L_{E1,\nu}^z$ and $L_{E2,\nu}^z$ carry no additional explicit factor of $\nu$ beyond the tilt.  Therefore a time-reversal pair with $v_{t,\nu}=\nu v_t$ obeys
\begin{equation}
L_{j,-\nu}^{z}=-L_{j,\nu}^{z},\quad j=1,2,3,
\label{eq:SM_TR_intrinsic_OAM_cancel}
\end{equation}
and its net intrinsic OAM vanishes.  In the conducting regime the occupations do change, and the next section shows that the regular cross-gap spin response reproduces a definite combination of the \emph{valley-resolved} $L_{E1,\nu}$ and $L_{E2,\nu}$ at quadratic order in Rashba coupling.  The insulating failure of conversion is therefore a valley-resolved statement for a time-reversal-symmetric pair; a net version requires valley imbalance or a time-reversal-breaking tilt pattern.

\section{Intrinsic Dirac spin response}

For the intrinsic $E_y\to s_z$ response in valley $\nu$, define the mixed spin--electric curvature of an exact conduction band $|+s,\nu\rangle$ at local momentum $\bq=k\hat{\bm x}$ by
\begin{equation}
{\cal A}_{sM,\nu}(k)=2\operatorname{Im}
\frac{\langle +s,\nu|s_z|M,\nu\rangle
\langle M,\nu|\partial_{q_y}H_\nu|+s,\nu\rangle}
{({\cal E}_s-E_M)^2}.
\label{eq:SM_A_spin}
\end{equation}
The sign follows from Eq.~\eqref{eq:SM_SE_intrinsic}; equivalently, the spin response is minus the occupied-band sum of the mixed curvatures.  Equation~\eqref{eq:SM_TR_spin_curvature} gives the paired-valley relation
\begin{equation}
{\cal A}_{sM,-\nu}(k,\phi+\pi)
=-{\cal A}_{sM,\nu}(k,\phi).
\label{eq:SM_A_spin_TR_Dirac}
\end{equation}
For the present rotational model the $zy$ component can be written
\begin{equation}
{\cal A}_{sM,\nu}(k,\phi)
={\cal A}_{sM}(k,0)\cos\phi,
\label{eq:SM_A_spin_rot}
\end{equation}
where the radial coefficient on the right is independent of $\nu$.  The minus sign in Eq.~\eqref{eq:SM_A_spin_TR_Dirac} is then supplied by $\phi\to\phi+\pi$.

Let $u_\nu=\hbar v_{t,\nu}$ and expand the scalar-tilt occupation at zero temperature:
\begin{equation}
f({\cal E}_s+u_\nu k\cos\phi)
=f({\cal E}_s)-u_\nu k\cos\phi\,\delta(\mu-{\cal E}_s)+O(u_\nu^2).
\end{equation}
Since $\int_0^{2\pi}\cos^2\phi\,\dd\phi=\pi$, the linear-tilt response of one valley is
\begin{align}
S_{z,\nu}^{\rm int}&=eE_yu_\nu K_{zy,\nu}^{\rm int},
\nonumber\\
K_{zy,\nu}^{\rm int}&=\sum_{s\in\mathrm{FS}}
\frac{k_s^2}{4\pi|\partial_k{\cal E}_s|_{k_s}}
\sum_{M\ne+s}{\cal A}_{sM,\nu}(k_s)
\equiv K_{zy}^{\rm int}.
\label{eq:SM_K_exact}
\end{align}
Thus $K_{zy,\nu}^{\rm int}$ is the same radial coefficient in the two valleys, while the physical response retains the explicit factor $u_\nu$.  For a time-reversal pair $u_{-\nu}=-u_\nu$ and therefore
\begin{equation}
S_{z,-\nu}^{\rm int}=-S_{z,\nu}^{\rm int},
\quad
\sum_{\nu=\pm}S_{z,\nu}^{\rm int}=0.
\label{eq:SM_TR_intrinsic_spin_cancel}
\end{equation}
This cancellation applies separately to the cross-gap and helicity sectors as long as the two valleys have equal occupations and the same dephasing parameters.

The terms with $M$ in the other conduction branch define $K^{\rm hel}$; the terms with $M$ in either valence branch define $K^{\rm cv}$.  The cross-gap term is regular as $\lambda_R\to0$. Expanding the exact eigenspinors, both Fermi contours, and their radial derivatives gives, in either valley,
\begin{equation}
K_{zy,\nu}^{\rm cv}=K_{zy}^{\rm cv}
=-\frac{\lambda_R^2(\mu^2+m^2)}
{4\pi\alpha^2\mu^4}+O(\lambda_R^4),
\quad \mu>m.
\label{eq:SM_Kcv_weak}
\end{equation}
For the same valley, the intrinsic OAM terms are
\begin{align}
L_{E1,\nu}^z
&=eE_yv_{t,\nu}\frac{m^2+3\mu^2}{48\pi\mu^3},
\\
L_{E2,\nu}^z
&=eE_yv_{t,\nu}\frac{m^2}{12\pi\mu^3},
\quad L_{E3,\nu}^z=0.
\label{eq:SM_L12}
\end{align}
The elementary identity
\begin{equation}
\frac{\mu^2+m^2}{4\pi\mu^3}
=4\frac{m^2+3\mu^2}{48\pi\mu^3}
+2\frac{m^2}{12\pi\mu^3}
\end{equation}
then proves the valley-resolved conversion relation
\begin{equation}
S_{z,\rm cv,\nu}^{\rm int,(2)}
=-\frac{\hbar\lambda_R^2}{\alpha^2\mu}
(4L_{E1,\nu}^z+2L_{E2,\nu}^z).
\label{eq:SM_L12_conversion}
\end{equation}
The $L_{E1}$ coefficient comes from the velocity-sum metric structure, while the two pieces of $L_{E2}$ combine into the second coefficient in Eq.~\eqref{eq:SM_L12_conversion}. This is an equality between the integrated conducting responses of a fixed valley, not merely a resemblance between connection products.  In a time-reversal-symmetric pair both sides are valley odd and sum to zero.

In the gap there is no Fermi contour, and a scalar tilt changes neither the exact projectors nor the insulating occupations. Hence, valley by valley,
\begin{equation}
S_{z,\nu}^{\rm int}=0,
\quad
L_{E1,\nu}^z=L_{E2,\nu}^z=\frac{eE_yv_{t,\nu}}{12\pi m}.
\label{eq:SM_gap_counterexample}
\end{equation}
The conducting conversion relation therefore cannot be continued through the band edge even at the level of a single valley.  For a time-reversal pair with $v_{t,\nu}=\nu v_t$, however, the two valley OAMs in Eq.~\eqref{eq:SM_gap_counterexample} cancel, so the \emph{net} insulating OAM is also zero.  The counterexample is therefore a valley-resolved failure of conversion in the time-reversal-symmetric model; it becomes a net finite-OAM/zero-spin example only if the valleys are unequally occupied or the tilt pattern itself breaks time reversal.

The helicity term has a different weak-coupling limit. The two individual Fermi-contour contributions diverge as $1/\lambda_R$, but their sum is finite in either valley:
\begin{equation}
K_{zy,\nu}^{\rm hel}=K_{zy}^{\rm hel}
=\frac{1}{4\pi\alpha^2}+O(\lambda_R^2).
\label{eq:SM_helicity_limit}
\end{equation}
This nonanalytic clean limit comes from the pair of exact conduction branches and has no parent-cross-gap OAM interpretation.  The associated spin density is still proportional to $u_\nu$ and therefore cancels between time-reversed valleys.

\enlargethispage{3\baselineskip}
For comparison with the Edelstein term, let $\bar\tau=\tau m/\hbar$ and $\bar u=\hbar v_t/\alpha$. With equal electric-field magnitudes,
\begin{equation}
\left|\frac{S_z^{\rm int}}{S_y^{\rm Ed}}\right|
=\frac{|uK_{zy}^{\rm int}|}{\tau|C_{yx}^{\rm Ed}|}.
\label{eq:SM_ratio}
\end{equation}
The exact Edelstein coefficient is evaluated from the same $4\times4$ eigenspinors, not from its weak-Rashba limit.

\setcounter{dbltopnumber}{3}
\renewcommand{\dbltopfraction}{1}
\renewcommand{\textfraction}{0}
\setlength{\dblfloatsep}{5pt}
\setlength{\dbltextfloatsep}{7pt}
\begin{table*}[!t]
\caption{Parameters for the fixed-field two-band TMDC calculation at $E_z=10^{-2}$ V/\AA.}
\label{tab:SM_tmdc_parameters}
\centering
\begin{tabular}{lrrrrrr}
\toprule
material & $a$ (\AA) & $t$ (eV) & $\Delta$ (eV) & $2\lambda$ (eV) & $|\lambda_{\rm BR}|$ (meV\,\AA) & $\lambda_R$ (meV)\\
\midrule
MoS$_2$  & 3.193 & 1.10 & 1.66 & 0.15 & 0.33 & 0.078\\
MoSe$_2$ & 3.313 & 0.94 & 1.47 & 0.18 & 0.55 & 0.130\\
WS$_2$   & 3.197 & 1.37 & 1.79 & 0.43 & 1.30 & 0.266\\
WSe$_2$  & 3.310 & 1.19 & 1.60 & 0.46 & 1.80 & 0.366\\
\bottomrule
\end{tabular}
\par\vspace{2pt}
\caption{Counterflow ratio ${\cal B}$ versus electron Fermi energy measured from the spinless conduction-band edge.}
\label{tab:SM_tmdc_doping}
\centering
\begingroup
\small
\setlength{\tabcolsep}{3pt}
\begin{tabular}{lrrr}
\toprule
material & 25 meV & 100 meV & 200 meV\\
\midrule
MoS$_2$  & 0.985 & 0.940 & 0.882\\
MoSe$_2$ & 0.983 & 0.932 & 0.867\\
WS$_2$   & 0.986 & 0.944 & 0.890\\
WSe$_2$  & 0.984 & 0.937 & 0.877\\
\bottomrule
\end{tabular}
\endgroup
\end{table*}

\section{Material-parameterized TMDC calculation}
\label{sec:SM_material_calculation}

The fixed-field TMDC calculation uses
\begin{align}
H_\nu={}&\alpha(\nu k_x\sigma_x+k_y\sigma_y)
+m\sigma_z
\nonumber\\
&-\nu\frac{\lambda}{2}(\sigma_z-1)s_z
+\lambda_R(\nu\sigma_xs_y-\sigma_ys_x),
\end{align}
with $\alpha=at$ and $m=\Delta/2$. The numerical parameters are listed in Table~\ref{tab:SM_tmdc_parameters}.
The $a,t,\Delta$, and $2\lambda$ values are the first-principles fits of Xiao \emph{et al.}, Phys. Rev. Lett. \textbf{108}, 196802 (2012). At fixed $E_z=10^{-2}$ V/\AA, Korm\'anyos \emph{et al.}, Phys. Rev. X \textbf{4}, 011034 (2014), give $|\lambda_{\rm BR}|=(0.033,0.055,0.130,0.180)E_z$ eV\,\AA. Matching $P_+H_RP_+=(\alpha\lambda_R/m)(k_xs_y-k_ys_x)+O(k^2)$ gives $\lambda_R=(m/\alpha)|\lambda_{\rm BR}|$. These unscreened values are upper estimates; their phase defines the measured in-plane spin axis. This matching constrains the projected conduction-band spin splitting, but it does not uniquely determine the off-diagonal completion $P_+H_RP_-$ or an embedding connection omitted by the two-band basis. The numbers below consequently test the minimal momentum-independent Rashba completion written above.

The parent conduction and valence bands in valley $\nu$ are defined by
\begin{equation}
H_{\orb,\nu}=\alpha(\nu k_x\sigma_x+k_y\sigma_y)+m\sigma_z.
\end{equation}
The Ising and Rashba terms obey Eq.~\eqref{eq:SM_TR_SOC}. Both are momentum independent, so $v_i=v_i^{\orb}=v_i^{\orb,\parallel}+v_i^{\orb,\perp}$, although both enter every exact energy and spinor. Each circular Fermi contour contributes
\begin{equation}
{\cal C}_{yx,\nu}^{X}
=\sum_N\frac{k_{F,\nu N}}{(2\pi)^2}
\int_0^{2\pi}\!\dd\phi\,
\frac{\langle s_y\rangle_{\nu N}
\langle\partial_{k_x}H^{X}\rangle_{\nu N}}
{|\partial_kE_{\nu N}|},
\end{equation}
where ${\cal C}=\hbar C$ and $X=(\orb,\parallel),(\orb,\perp)$. Time reversal gives ${\cal C}_{yx,-\nu}^{X}={\cal C}_{yx,\nu}^{X}$, so Table~\ref{tab:SM_tmdc_doping} sums two equally occupied valleys and gives the doping dependence of the counterflow ratio.
Counterflow remains $87$--$99\%$ over $E_F=25$--$200$ meV. Halving $E_z$ changes ${\cal B}$ by less than $10^{-4}$ and ${\cal C}/E_z$ by less than $0.2\%$ at $E_F=100$ meV, confirming the linear-field regime. Additive closure is better than $6\times10^{-16}$ nm$^{-1}$. The Pauli-spin polarisation is $S_y=(e\tau E_x/\hbar){\cal C}_{yx}$; multiplication by $\hbar/2$ gives spin-angular-momentum density.

\end{document}